\documentclass[10pt, conference, letterpaper,onecolumn]{IEEEtranv1.8}

\usepackage{graphicx}  
\usepackage{subcaption}
\usepackage{amsmath}  
\usepackage{amsfonts}
\usepackage{balance}
\usepackage{amsthm}
\usepackage{algorithmic}
\usepackage{algorithm}
\usepackage{xspace}
\usepackage{booktabs}
\usepackage[noadjust]{cite}

\usepackage[colorlinks]{hyperref}
\hypersetup{
    bookmarks=true,         % show bookmarks bar?
    unicode=false,          % non-Latin characters in AcrobatÕs bookmarks
    pdftoolbar=true,        % show AcrobatÕs toolbar?
    pdfmenubar=true,        % show AcrobatÕs menu?
    pdffitwindow=false,     % window fit to page when opened
    pdfstartview={FitH},    % fits the width of the page to the window
    pdftitle={Hashing-Pursuit for Anomaly Detection}, % title
    pdfauthor={anon, et al.},     % author
    pdfsubject={Network Security},   % subject of the document
    pdfcreator={Anonymous},   % creator of the document
    pdfproducer={}, % producer of the document
    pdfkeywords={networks} {anomaly} {ddos} {detection} {hashing}, % list of keywords
    pdfnewwindow=true,      % links in new window
    colorlinks=true,       % false: boxed links; true: colored links
    linkcolor=red,          % color of internal links
    citecolor=blue,        % color of links to bibliography
    filecolor=magenta,      % color of file links
    urlcolor=cyan           % color of external links
}

\renewcommand\P{\mathbb P}
\newcommand\E{\mathbb E}

\theoremstyle{plain}% default
\newtheorem{thm}{Theorem}%[section]

\newtheorem{cor}{Corollary}
\newtheorem{fact}{Fact}
\theoremstyle{definition}
\theoremstyle{remark}
\newtheorem{rem}{Remark}

\def\HideProof#1{}

\begin{document}

\title{Hashing Pursuit for Online Identification of Heavy-Hitters in High-Speed Network Streams}

\author{
 \IEEEauthorblockN{
  Michael G. Kallitsis\,\IEEEauthorrefmark{2},
  Stilian Stoev\,\IEEEauthorrefmark{1},
 and George Michailidis\,\IEEEauthorrefmark{1} 
 }
 \IEEEauthorblockA{\IEEEauthorrefmark{1}\,Department of Statistics, University of Michigan, Ann Arbor, MI\\}
  \IEEEauthorblockA{\IEEEauthorrefmark{2}\,Merit Network, Inc., Ann Arbor, MI\\}
 \IEEEauthorblockA{\{mgkallit, sstoev, gmichail\}@umich.edu}}

\maketitle

\begin{abstract}
Distributed Denial of Service (DDoS) attacks have become more prominent recently, both in frequency of occurrence, as well as magnitude. Such attacks render key Internet resources unavailable and disrupt its normal operation. It is therefore of paramount importance to quickly identify malicious Internet activity. The DDoS threat model includes characteristics such as: (i) heavy-hitters that transmit large volumes of traffic towards ``victims", (ii) persistent-hitters that send traffic, not necessarily large, to specific destinations to be used as attack facilitators, (iii) host and port scanning for compiling lists of un-secure servers to be used as attack amplifiers, etc. This conglomeration of problems motivates the development of space/time efficient summaries of data traffic streams that can be used to identify heavy-hitters associated with the above attack vectors. This paper presents a hashing-based framework and fast algorithms that take into account the large-dimensionality of the incoming network stream and can be employed to quickly identify the culprits. The algorithms and data structures proposed provide a synopsis of the network stream that is not taxing to fast-memory, and can be efficiently implemented in hardware due to simple bit-wise operations. The methods are evaluated using real-world Internet data from a large academic network.
\end{abstract}

\section{Introduction}

Distributed Denial of Service attacks have become prominent recently both in frequency of occurrence as
well as magnitude~\cite{rossow}.  %~\cite{rossow, prolexic2014q1, ars_jan_gamer_attacks}. 
The detection and identification of such nefarious Internet activity are key problems for network engineers. The {\em time scale} at which such attacks 
are detected and {\em identified}
is of crucial importance. In practice, this time scale, referred here as the {\em relevant time scale} (RTS),
should be on the order of {\em seconds} or {\em minutes} rather than hours. 
However, processing data in a streaming fashion for network monitoring
that aims to detect and identify network anomalies
poses two fundamental computing challenges. First, one needs to work
with ``small-space" data structures;  storing snapshots of the incoming data stream  in fast memory
 is prohibitively expensive. 
Second, any data processing  on the incoming network stream 
ought to be performed efficiently; expensive and time-consuming
computations on voluminous streams may defeat the purpose of
real or near-real time network monitoring.

In addition to the need for rapid RTS detection and identification of malicious activities, another important feature is their growing
sophistication.  Attacks in complex modern networks are often distributed and coordinated~\cite{rossow}. For example, the sources involved in a 
DDoS attack may be spread through various sub-networks and individually may not stand out as heavy traffic generators. Therefore, 
detection of the attack and identification of its victim(s) is only possible if traffic is monitored  {\em simultaneously} over multiple sites on 
the network. The rapid communication (on the RTS scale) of anomaly signatures from monitoring sites to a single decision center imposes stringent 
size constraints on the data structures involved.

%The key contributions of this paper are as follows: We introduce fast and
%space-efficient algorithms based on \emph{hashing}. 

%We propose
%computationally efficient hash functions that 
%help us construct our summary data structures while at the same time ``spreading" and
%minimizing collisions that are, of course, inevitable when hash arrays are used.
%The proposed algorithms are space and time efficient and help identify
%the heavy-hitters with remarkable accuracy. 

In  recent years, numerous sophisticated and accurate methods for anomaly detection have been proposed. For example,
\emph{signature-based} methods examine traffic/data for
\emph{known} malicious patterns to detect malware (worms, botnets, etc); \emph{Snort} (see, {\tt snort.org})  and \emph{Bro} 
({\tt bro.org}) are two well-known tools of this class.
On the other hand, 
a plethora of methods have appeared in the literature that look for deviations from ``normality" (e.g., see~\cite{Lakhina:2004:DNT:1030194.1015492, Barford:2002:SAN:637201.637210, Ide04eigenspace-basedanomaly, Zhang:2005:LAD:1080173.1080189}
to name a few, and the review paper~\cite{Thottan:2010:anmomaly:detection:review} for a more exhaustive list)
and do not require a priori knowledge of attack patterns.
Most of them, however, 
require heavy computation (e.g., singular value decomposition of large matrices or computation of wavelet coefficients) and/or storage and post-processing of historical 
traffic traces. This makes them essentially not applicable in practice on the RTS. 
%Further, as pointed out by~\cite{Thottan:2010:anmomaly:detection:review}, though, fine calibration of PCA/subspace construction methods~\cite{Lakhina:2004:DNT:%1030194.1015492, Lakhina:2004:SAN:1005686.1005697, Lakhina:2005:MAU:1080091.1080118}
%could be challenging. For example, the model can be sensitive to the numbers of principal components used
%to construct the normal subspace.  
%On the other hand, the rapidly 
%RTS--computable traffic summaries available to network operators are often crude and lack the ability to precisely identify individual 
%IP addresses. 
%Finally, PCA methods are inferior to wavelet methods when it comes to detecting Internet scanning~\cite{5934975}.
%The general methodology and specific algorithms proposed in our paper aim to bridge this gap.
As a consequence, much attention was paid to alleviating the high dimensionality constraint of the problem. Algorithms that involve
memory efficient data structures, called ``sketches", 
that exhibit sub-linear computational efficiency  with respect to the input space
have been well-studied by the community~\cite{Cormode:2005:IDS:1073713.1073718, Cormode:2003:FHH:1315451.1315492, Cormode:2004:HUS:1007568.1007575, Gilbert06algorithmiclinear, Krishnamurthy:2003:SCD:948205.948236, 4146856, 1354567, 2190032, Cormode1061325,  Porat:2012:STM:2095116.2095212, doi:10.1137/100816705}. These summary data structures
can accurately \emph{estimate} the 
state of the measured signal while having a low computation and space fingerprint.
However, with the  exception of the work in~\cite{4146856}, where a
hash-based algorithm and  fast 
hardware implementation is introduced, 
there is a  gap between the theoretical optimality of advanced sub-linear algorithms
and their practical implementation.

This paper aims to bridge this gap and its key {\em contributions} include:
(i) the  development of space and time efficient \emph{algorithms  and data structures} that can be continuously 
and rapidly updated while monitoring fast traffic streams. The proposed data structures, based on permutation hashes, allow us to identify the
heavy-hitters in a traffic stream (i.e., the IPs or other \emph{keys} responsible for the, say, top-$k$ signal values in a window of interest).
Our algorithms are
memory efficient and require constant memory space that is much smaller than the 
dimension of the streaming data. Further, all computations
involve fast bit-wise  operations making them amenable for fast hardware implementation;
(ii) we propose a \emph{framework  suited for different types} of traffic signals in a unified ``hashing pursuit" perspective. For example, the signal of interest can be 
conventional traffic volume (measured in packets or bytes) or the number of different source IPs  that have accessed a given destination, etc. In 
the latter case (upon filtering out the well-known benign nodes/users/networks) the heavy hitters correspond to potential victims of DDoS attacks. Their rapid
identification (on the RTS) for the purpose of mitigation is of utmost importance to network security; (iii) we evaluate our algorithms with
\emph{real-world networking data collected at a large academic ISP} in the United States.  
 
\section{Streaming Paradigm}
\label{sec:streaming}

The theoretical framework of computation on data streams is best suited to our needs~\cite{Muthukrishnan:2005:DSA:1166409.1166410}. In this context,
a traffic link can be viewed as a stream of items $(\omega_i,v_i),\ i=1,2,\ldots$ that are seen only once
as they pass through a monitoring station (e.g.\ traffic router). Due to space/time constraints, the 
entire stream cannot be recorded and hence only fast small memory sequential algorithms can be used 
for its characterization and analysis. The $\omega_i\in \Omega$ are the {\em keys} and $v_i$ are the updates 
(e.g., payload increments) of the stream. For example, the set of keys $\Omega := \{0,1\}^{32}$ could be all IPv4 addresses of
traffic sources and the payloads could include byte content, packets, port, protocol,
or other pertinent information. Alternatively, the stream keys $\omega = (s,d)\in \Omega := \{0,1\}^{64}$ could be the pair of
source and destination IPv4 addresses, and one could enlist many other types of keys (e.g., IPv6 addresses). 

Over a monitoring period (e.g.\ a few seconds), the stream communicates a {\em signal} $f:\Omega \to V$, which could
take either {\em numerical} or {\em set} values. Consider the simple scalar setting where $\omega_i = (s_i,d_i) \in \Omega \equiv \{0,1\}^{64}$ and
$v_i$ are the number of bytes exchanged between source $s_i$ and destination $d_i$. Upon observing the $i$-th update $(\omega_i,v_i)$, 
the signal $f$ is updated sequentially like a {\em cash register}\footnote{We used a programmers' syntax for  variable updates.}:
$$
 f(\omega_i) := f(\omega_i) + v_i.
$$
Thus, at the end of the monitoring period,  $f$ contains the number of bytes communicated for all pairs $\omega=(s,d)$. 
The ``signal", however, is only a theoretical quantity, since in practice, it cannot be fully stored and accessed on the RTS. Nevertheless, compressed representations of $f$, 
known as {\em sketches} have been proposed to study its characteristics~\cite{Cormode:2005:IDS:1073713.1073718, Cormode:2003:FHH:1315451.1315492, Cormode:2004:HUS:1007568.1007575, Gilbert06algorithmiclinear, Krishnamurthy:2003:SCD:948205.948236, 4146856, 1354567, 2190032, Cormode1061325,  Porat:2012:STM:2095116.2095212, doi:10.1137/100816705}. Broadly speaking sketches 
provide statistical summaries of the signal, which can be used to approximate (with high probability) features of interest.  (See also 
the related {\em compressed sensing} paradigm~\cite{4385788, Gilbert:2007:OSF:1250790.1250824, 1614066, Indyk:2008:ECC:1347082.1347086}.)

The cash register model described above is well-suited for applications such as  
monitoring IP traffic entering a network link, monitoring IPs accessing a particular service (such as Web server, cloud storage, etc), monitoring 
particular IP pairs of a particular application (i.e., source-destination pairs of all DNS traffic), and many others. 

A variation of the cash register model is obtained when the payloads $v_i$'s are sequential updates
on a set. In this model, upon observing $(\omega_i,v_i)$, we update the state of the signal as follows:
$f(\omega_i) := f(\omega_i) \cup \{v_i\}$. For example, if $\omega_i \in \Omega =\{0,1\}^{32}$ is the source IP and
$v_i\in \{1,2,\cdots,M\}$ is port number ($M=2^{16}$)
(we define the compact notation $[M]=\{1,2,\ldots,M\}$ to be used hereafter), then the signal $f:\Omega \to 2^{[M]}$, where $ 2^{[M]}$ is the power-set of $[M]$,  is set-valued.
For a given IP $\omega$, the set $f(\omega)$ consists of all different ports accessed by $\omega$ during the monitoring  period and
large values of ${\rm card}(f(\omega))$ identify potential {\em port scanning} sources $\omega$.
Similarly, by considering destination IPs in place of ports we can identify hosts that perform
malicious activity such as horizontal \emph{host scanning}.

In the above two contexts, the goal is to identify heavy--hitters, i.e.\ source--destination pairs generating heavy traffic, or a source 
with a large set of accessed ports or destinations, for example.  For a signal $f:\Omega\to V$, a heavy--hitter is formally
a key 
$$
\omega^*  = {\rm Argmax}_{\omega\in \Omega} {\cal L}(f(\omega)),
$$
maximizing a {\em loss} function ${\cal L}(f(\omega))$, where ${\cal L}$ could be a simple byte-count, size of a set, or 
in a more complex, time--series setting, the frequency maximizing the power spectrum, for example.

Our goal is to develop a unified framework for the identification of heavy hitters, which can be applied to a variety of signals and
that works on the {\em relevant time scale} for anomaly detection (i.e., the algorithms monitor the stream in real time 
and produce heavy--hitters every few seconds, minutes or every 100,000 tuples seen, etc). 
Depending on the type of signal, the identification part is bundled with a different specialized {\em sketch}, which 
allows us to handle ultra high dimensional signals. The following section describes the general methodology in terms of
meta-algorithms and then proceed to several concrete applications. 

\section{Hashing Pursuit for Identification}
\label{sec:hash-ident}

At the heart of our identification schemes lies \emph{hashing}.  Reversible hash functions are used to compress the 
\emph{domain} of our incoming signal into a smaller dimension, while at the same time uniformly distributing the 
keys onto the reduced-dimension space. Another application of hashing used below is to efficiently `thin' the 
original stream into sub-streams to help increase identification accuracy.
We apply our hash functions on a set of keys $\Omega = \{ \omega_1,\ldots, \omega_N\}$, important 
special case being the set of all IPv4 numbers, where $N= |\Omega| =2^{32}$.

%\subsection
\smallskip
\noindent{\bf Hash functions for IP monitoring.}
Consider a set of $q$
hash functions $h_i:\Omega \to [m],\ i \in [q]$. This set is {\em complete} (or perfect),
if all $\omega\in\Omega$ are uniquely identified by their hashes $h_i(\omega)$, $i\in[q]$, i.e.,
the vector--hash $H(\omega) = (h_i(\omega))_{i=1}^q$ is invertible.  We shall describe below a particular family of
rapidly computable complete hashes, which can also be quickly inverted. In particular, we will have that
$|\Omega| = m^q$.

We describe next the main idea behind identification over the simple case of a scalar signal $f:\Omega\equiv \{0,1\}^{32} \to [M]$,
which tracks the number of packets $f(\omega)$ transmitted by IP $\omega$ over a monitoring period. The goal is to be able to identify
the most {\em persistent user} with highest packet count.

Instead of maintaining in memory the entire signal $f$ using an array of size $|\Omega|$, we shall maintain $q$ hash--histograms 
$\tilde f_i,\ i \in[q]$, of size $m$ each. If $\omega^*$ is the heavy hitter, then it will contribute large packet counts
$\tilde f_i(o_i^*)$ to all of the $q$ histograms at the bins $o_i^* = h_i(\omega^*),\  i\in[q]$.
If collisions are uniformly spread--out, with high probability, we will have that
$$
o_i^* = \tilde o_i^* := {\rm Argmax}_{o_i\in[m]} \tilde f_i(o_i),\ \ i\in[q]
$$
Thus, by locating the bins $\tilde o_i^*$ maximizing each of the $q$ hash--histograms, we define
\begin{equation}\label{e:inverse}
\tilde\omega^* := \text{inverse}(\tilde o_1^*,\ldots,\tilde o_q^*).
\end{equation}
We provide lower bounds on $\P(\omega^* = \tilde \omega^*)$ and show that
in practice the heavy hitter is identified w.h.p. (with high probability).
In fact, the method naturally extends to the case of multiple heavy hitters and other 
types of signals. {\em Meta-algorithms}~\ref{alg:meta-encode} and~\ref{alg:meta-decode} 
describe the general encoding  and decoding steps. 

\renewcommand{\thealgorithm}{\roman{algorithm}}

\floatname{algorithm}{Meta-Algorithm}
\begin{algorithm}[t]
\caption{Hash-based Encoding}
\label{alg:meta-encode}
%\scriptsize
\begin{algorithmic}[1]
\REQUIRE Set of complete uniform hashes 
 $$
  h_i:\Omega\to \{1,\cdots,m\},\ i \in[q],
 $$
 with $|\Omega| = m^q$. Initialize signal hash histograms: 
 $$
  \tilde f_{i} = {\mathbf{zeros}}(1,m),\ i\in[q].
 $$

\STATE [Start] Begin stream monitoring.
\STATE [Hash keys] Upon observing $(\omega,v)$, compute $o_i = h_i[\omega],\ i\in[q]$.
\STATE [Update] Compute $ \tilde f_i[o_i]$ = ${\rm update}( \tilde f_i[o_i], v)$,  $i\in[q]$
\STATE [Stop] End when the monitoring period is over.
\RETURN Output hash arrays $\tilde f_i,\ i\in[q]$ for analysis.
\end{algorithmic}
%\end{algorithm}
\end{algorithm}

\begin{algorithm}[t]
\caption{Decode and Identify}
\label{alg:meta-decode}
%\scriptsize
\begin{algorithmic}[1]
\REQUIRE Set of hash histogram arrays $\tilde f_i,\ i\in [q]$, as outputs of
Meta-algorithm \ref{alg:meta-encode}. An efficient ${\rm inverse}$ function as in \eqref{e:inverse}.

\STATE [Identify] For each hash histogram array $\tilde f_i,$ identify a bin index $o_i,\ i\in [q]$ as a
candidate bin where the heavy hitter falls.
 
\RETURN $\omega = {\rm inverse}(o_1,\ldots,o_q)$.
\end{algorithmic}
\end{algorithm}

\renewcommand{\thealgorithm}{\arabic{algorithm}}

The specific implementation of the hash functions and their inverses will be discussed in the following section, where
the abstract {\rm `update'} and {\rm `identify'} steps in the above meta--algorithms will depend on the application. 
%We outlined the encoding and decoding steps in terms of meta algorithms in order to outline the general framework.

\begin{rem} Let $|\Omega| = m^q$. The space required to maintain the raw signal $f:\Omega\to V$, is
${\cal O}(m^q)$, while the space required for the hash histograms $\tilde f_i,\ i\in[q]$ is ${\cal O}(m\times q)$. These
exponential savings in memory allow us in practice to rapidly communicate hash--based traffic summaries at the RTS, 
enabling simultaneous and distributed monitoring and identification (see Table~\ref{tab:mem}).
\end{rem}

\begin{rem}\label{rem:update} The update step in Algorithm \ref{alg:meta-encode} depends on the type of signal. In the scalar case, we have
$$
{\rm update}( \tilde f_i[o_i], v) = \tilde f_i[o_i] + v,
$$
while in the set-valued case ${\rm update}( \tilde f_i[o_i], v) = \tilde f_i[o_i] \cup\{ v\}$.
In applications, maintaining a {\em set}, however, is often not practical either. When we are only interested in the size of the set, 
we can maintain a further max--stable sketch data structure, requiring an array of size $L$ for each $o_i\in [m],\ 
i\in[q]$ (see Section~\ref{sec:ms}).
\end{rem}

%\subsection{Construction of complete reversible hashes}
\smallskip
\noindent{\bf Construction of complete reversible hashes.}
We focus on the case where the keys represent IPv4 numbers, i.e.\ $\Omega = \{0,1\}^{32}$. A simple set of complete
hash functions can be constructed as follows. Let $m = 2^p := 2^8$ and $q = 4$ and consider the base-$m$ expansion of $\omega$:
$$
\omega = \overline{o_q\cdots o_1} = \sum_{i=0}^{q-1} o_{i+1} m^i.
$$
The $o_i$'s are simply the 4 \emph{octets} in the dotted-quad representation of the IP address $\omega\in \Omega$. 

Clearly, the hashes 
\begin{equation}\label{e:hi}
h_i(\omega):= o_i,\ i\in [q]
\end{equation}
are rapidly computable with bit--wise operations, e.g.\ $o_i  = {\tt ShiftRight}(\omega, (i-1)*8)\ {\rm mod}\ 256$.  They are also complete since
the knowledge of all $q$ base-$m$ digits determines $\omega$. In practice, however, these naive hash functions are not useful. The reason
is that the IPs are typically clustered (spatially correlated) since the address space used by a sub--network or \emph{autonomous system}
consists of contiguous ranges of IPs. Hence, one needs to permute or `mangle' the IPs
in the original space $\Omega$ so that all permuted keys achieve approximately uniform empirical distributions.
This can be readily resolved by composing the hashes with a permutation that can be also efficiently 
computed {\em and} inverted.  The idea of IP mangling was employed also in~\cite{4146856}. 
Next, we describe a different kind of permutation functions, starting first with a simple general result.

\begin{fact}  Equip the set of keys $\Omega = \{0,1\}^{p\times q}$ with the uniform probability measure. Then:

 {\rm (i)} The hashes $h_i, i\in [q]$ in \eqref{e:hi}, viewed as random variables on $\Omega$,
  are independent and uniformly distributed.
 
 {\rm (ii)} For any permutation $\sigma:\Omega\to\Omega$, the hashes $h_i\circ \sigma,\ i\in[q]$ are also independent
 and uniformly distributed.
 
Conversely, if $\tilde h_i:\Omega \to [2^p],\ i\in[q]$ is a set of uniformly distributed independent hashes, then they are complete and
$\tilde h_i = h_i\circ \sigma$, for some permutation $\sigma$.
\end{fact}

\noindent The proof is elementary and omitted in the interest of space. This fact suggests that so long as a permutation $\sigma$ spreads-out 
a set of IPs essentially uniformly, the empirical distribution of the hashes $h_i\circ \sigma$ will be approximately uniform (see Fig.~\ref{fig:ident_typical}).
Therefore, to obtain a high--quality set of complete hashes, it suffices to find an efficiently computable and invertible permutation $\sigma$. 
We do so next.

\begin{fact} \label{f:RSA} Let $\Omega = [N]$ and suppose that $\sigma_0 \times \sigma_1 = 1 \ (\rm{mod }\ N)$, for some integers 
$\sigma_0,\ \sigma_1\ge 2$. Define $\sigma, \tilde \sigma :\Omega \to \Omega$, as follows
$$
\sigma (\omega) := \omega \times \sigma_0 \ (\rm{mod }\ N)
\ \mbox{ and }\ 
\tilde\sigma (\omega) := \omega \times \sigma_1 \ (\rm{mod }\ N). 
$$
Then,  $\sigma$ and $\tilde \sigma$ are bijections and $\sigma^{-1} = \tilde \sigma$.
\end{fact}

\noindent 
The proof is straightforward. 

\begin{rem}Similar but more computationally intensive IP mangling methods (with efficient hardware 
implementation on an FPGA) were given in \cite{4146856}. Shai Halevi \cite{halevi:2007} proposed 
alternative permutations that can be efficiently computed and inverted and exhibit good cryptographic
properties. Interesting connections with the general theory of {\em homomorphic coding} can be further pursued. 
\end{rem}

%\begin{proof}
% %Since $\sigma_0$ and $\sigma_1$ divide $N+1$, it follows that $\sigma_0$ and $\sigma_1$ are co--prime with $N$. Thus if
% %$\omega\times\sigma_0 = \omega' \times \sigma_0 (\mod N)$, we have $(\omega - \omega')\times \sigma_0$ is divisible by $N$, which means that
% %$\omega  = \omega' (\mod N)$, and hence $\omega = \omega'$. This proves that $\sigma$ and $\tilde \sigma$ are one-to-one. We will show that $\sigma \circ \tilde \sigma ={\rm id}$,
% %which would then entail that $\sigma$ are bi
%
%We will show that $\sigma \circ \tilde \sigma = \tilde \sigma \circ \sigma = 1$, which would automatically imply both claims. For every
%$\omega\in [N]$, we have $\sigma \circ \tilde \sigma (\omega) = \omega \times \sigma_0 \sigma_1  \ (\rm{modulo }\ N) $. The fact that $\sigma_0 \sigma_1 \equiv 1  \ (\rm{modulo }\ N)$ 
%implies that for some integer $k$, we have $\omega \times \sigma_0 \sigma_1 = \omega + k N = \omega   \ (\rm{modulo }\ N) $. This shows $\sigma \circ \tilde \sigma = 1.$ 
%Note that $\sigma$ and $\tilde \sigma$ commute by definition.
%\end{proof}

In practice, with $\Omega=\{0,1\}^{32}$, we consider 
\begin{equation}
\label{e:2-32-factorization}
   2^{32} +1 = 641 \times 6,700,417 =: \sigma_0 \times \sigma_1.
\end{equation}
We found that $h_i \circ \sigma$ with $\sigma_0 = 641$ works well. Observe also that if $\sigma_0 \sigma_1 = 1 \mod N$, then $\sigma_0^\gamma \sigma_1^\gamma = 1 \mod N$, 
for all integer $\gamma$. Thus, via powers of $\sigma_0$ and $\sigma_1$ in \eqref{e:2-32-factorization}, one can easily generate different permutations. 
Our experiments with naturally occurring sets of IPs show powers $\gamma=3$ or $5$ work slightly better than $\gamma=1$.

\section{Identification Algorithms}
\label{sec:identification}

%The IP mangling step introduced above translates into good ``uniformization" of the empirical distributions of hashes
%and therefore removes any correlation or spatial locality of the input stream (see also~\cite{Schweller:2004:RSE:1028788.1028814}).
%At the same time, one can ``reverse" the permutation function and recover the malicious IPs in the original input space.
Having all  building blocks in place, we now introduce our algorithms. Table~\ref{tab:roadmap} provides a roadmap.

\begin{table}[t]
\caption{Algorithms Roadmap}
\scriptsize
\centering
\label{tab:roadmap}
\begin{tabular}{|l|l|c|}
\hline
{\bf Algorithm}                & {\bf Sketch-size}             & {\bf Application}  \\   
\hline\hline                                                              
Simple~(\ref{sec:simple})                   & $O(q\times m$)          & Scalar signals, top-1 hitter, suited  \\
                    &           &                            for stringent memory constraints  \\
\hline                    
Max-Count~(\ref{sec:max_count})                & $O(q\times m \times m'$) & Scalar signals, top-k hitters                                                \\
\hline
Boyer-Moore~(\ref{sec:bm}) & $O(m\times m'$)          & Scalar signals, top-k hitters                                                \\
\hline
Max-Sketch~(\ref{sec:ms})               & $O(q\times m \times L$) & Set-valued signals, top-hitter         \\
\hline                                     
\end{tabular}
\end{table}

\subsection{Simple Hashing Pursuit}
\label{sec:simple}

%Let ${\cal S} = \{ (j ,I_i),\ i=1,2,\ldots\}$ be a data stream, where the users are identified with their IPv4 keys $j \in \Omega$. We shall 
%assume that {\em one} user $\omega_0$ generates abnormally large traffic and the goal is to efficiently identify it without having to store the stream.

%\medskip
%\noindent{\bf Algorithm.} ({\bf Simple Hashing Pursuit})
%\begin{enumerate}
%   \item {\em (initialization)} Initialize array $P$ of size $q \times m$. That is, set
%    $P[i, j] := 0$ for $i \in [q]$  and  $j\in [m]$.
% 
%   \item {\em (permute)} Upon observing $(\omega, v)$, compute $\sigma (\omega)$.
%   
%   \item {\em (hashing)} After IP mangling, compute $h_i\circ \sigma (\omega), \ i \in [q]$.
%   
%   \item {\em (signal updates)} Update the $q$ signal arrays of size $m$ as
%   $
%    P[i, h_i \circ \sigma(\omega)] = P[i,h_i\circ \sigma(\omega)] + v.
%   $
%   
%   \item {\em (identification oh heavy-hitter)} 
%   \begin{enumerate}
%   \item {\em (select)} Upon populating the signal arrays for some period of time,
%   find
%   $$
%    o_i := {\rm Argmax}_{j=1\in[m]} P[i,j],\ \ i \in [q].
%   $$
%  \item (\emph{decode})
%   Output $\omega^* := \sigma^{-1}(\overline{o_4 \cdots o_1})$
%   \end{enumerate}
%\end{enumerate}

\floatname{algorithm}{Algorithm}
\setcounter{algorithm}{0}
\begin{algorithm}[t]
\caption{Simple Hashing Pursuit (SHP)}
\label{alg:shp}
%\scriptsize
\begin{algorithmic}[1]
\REQUIRE Number heavy-hitters: $k$
\STATE [Start]   Initialize array $P$ of size $q \times m$.
\STATE [Permute]  Upon observing $(\omega, v)$, compute $\sigma (\omega)$
\STATE [Hash]  Compute $h_i\circ \sigma (\omega), \ i \in [q]$.  \# NOTE: $f\circ g (x):=f(g(x))$ 
\STATE [Update]    Update  $P$ for $i \in [q]$:
   $$
    P[i, h_i \circ \sigma(\omega)] = P[i,h_i\circ \sigma(\omega)] + v.
   $$
 \vspace{-10pt}  
\IF {(Populated $P$ for some  time window)}
\IF{k==1}
 \STATE Find $o_i = {\rm Argmax}_{j} P[i,j]$, \ $i\in[q]$ 
 \STATE [decode] Output $\omega^* = \sigma^{-1}(\overline{o_q \ldots o_1})$. 
 \ELSE
  \STATE Initialize ${\cal O}_i = \emptyset$, \ $i\in[q]$ 
   \FOR{r=1 \TO k} 
        \STATE Find $o_i = {\rm Argmax}_{j\in[m] \setminus {\cal O}_i  } P[i,j]$, \ $i\in[q]$ 
        \STATE ${\cal O}_i  = {\cal O}_i \cup  o_i $  {\#Exclude for next iteration }
         \STATE [decode] Output $\omega_r^* = \sigma^{-1}(\overline{o_q \ldots o_1})$. 
   \ENDFOR
 \ENDIF    
 \ENDIF    
\end{algorithmic}
%\end{algorithm}
\end{algorithm}

%\medskip
%\noindent\emph{Parameter choices:} For identification of hosts in IPv4 where
%$N=2^{32}$, we suggest setting $q=4$ and $2^p = 256.$ \TODO{For pairs and IPv6...}

%\fbox{ Refer to the chaining pursuit by Gilbert et al}.

\begin{figure}
        \centering
        \includegraphics[width=0.34\textwidth]{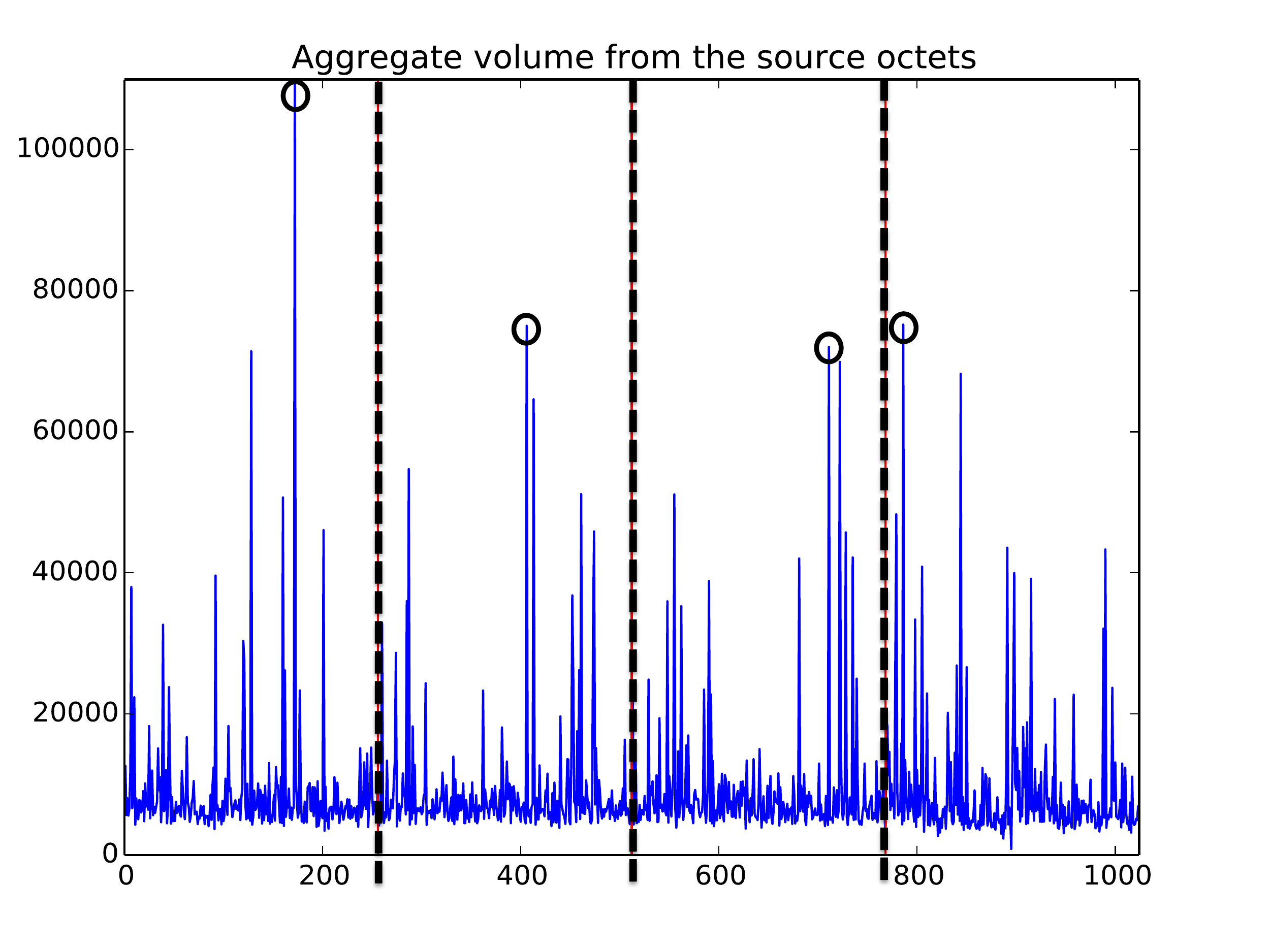}
        \caption{Algorithm~\ref{alg:shp}'s data structure consists of $q=4$ hash arrays of size $m=256$ (here separated by the thick-dashed lines), one for each octet. Circles illustrate the selected maxima of each octet.}
        \label{fig:ident_typical}
\end{figure}

Algorithm~\ref{alg:shp} is the simplest instance of IP randomization and hashing used to identify the
heavy hitter for a scalar signal.  Fig.~\ref{fig:ident_typical} shows the resulting {\em signal arrays} 
$P[i,j],\ j \in [256],\ i\in[4]$, referred also as \emph{octet arrays}. If $o_1$ is the first octet of 
the (encoded) IP address $\omega$, then its payload $p$ is added to the bucket $P[1,o_1]$.
If the permutation $\sigma$ spreads out the range of IPs in the observed stream approximately uniformly,
 it is reasonable to expect that the resulting vector $P[1,j],\ j\in[256]$ 
will be populated uniformly.  A heavy hitter, however, will contribute abnormally large signal value to the bucket
$P[1,o_1^*]$, where  $o_1^*$ is the first octet of its encoded key. The heavy hitter similarly stands out in its encoded octet indexes $o_2^*, o_3^*$ and $o_4^*$. By applying the inverse of the permutation $\sigma$ to the largest bins 
in the octet arrays, we recover the address of the anomalous user (see also Meta-Alg.\ \ref{alg:meta-decode}). 

The fact that the permutation $\sigma$ can be efficiently inverted is of utmost importance in practice 
since this algorithm is intended to be run online on fast network traffic streams. This is, in fact, feasible on the RTS
as demonstrated in~\cite{4146856} with an efficient FPGA implementation of a similar and slightly more 
computationally demanding IP randomization.

%\begin{figure}
%        \centering
%        \begin{subfigure}[b]{0.45\textwidth}
%                \includegraphics[width=\textwidth]{./figures/Fig_ident_alg_example.pdf}
%                \caption{A typical heavy hitter (CHI-710a, 2014-01-29 12pm)}
%                \label{fig:ident_typical}
%        \end{subfigure}
%        \quad
%        \quad
%        \begin{subfigure}[b]{0.45\textwidth}
%                \includegraphics[width=\textwidth]{./figures/Fig_ident_alg_example_extreme.pdf}
%                \caption{An extreme heavy hitter (WSUb, 2014-01-29 12pm)}
%                \label{fig:ident_heavy}
%        \end{subfigure}
%        \caption{\TODO{Need Stilian's original figures.}}
%\end{figure}

\subsection{Hashing Pursuit on Sub-streams}
\label{sec:max_count}

Algorithm~\ref{alg:max-count}, called~\emph{Max-Count Hashing Pursuit}, is an extension of SHP that minimizes collisions significantly, 
and hence improves identification accuracy. The idea is to introduce an
additional dimension $m'$ in our data structure and populate it by using another hash function $h_t$. The secondary
hash is used to divide the incoming stream into $m'$ independent sub-streams (i.e., `thin' the stream). 

%First, we discuss how the sub-streams should be constructed.
%Given $p \in (0,1)$, we want to generate a sub-stream ${\cal S}_p = \{ (j,  I_i),\ j \in  \Omega$\}, such that
%for each $j \in {\cal S}$, the point $j \in {\cal S}_p$ with probability $p$, regardless of the frequency that key $j$ appears
%in the stream ${\cal S}$.
%
As stream entries arrive, we initially perform the \emph{permutation}
step discussed above to remove spatial localities.
Then, having a uniform hash function
$h_t: [N] \to [m']$ as discussed in Section~\ref{sec:hash-ident},
we apply it on the permuted key $\omega$ to get the index $s$ of the sub-stream.
We update the appropriate array entries corresponding to sub-stream $s$ only.

%\medskip
%\noindent{\bf Algorithm.} ({\bf Max-Count Hashing Pursuit})
%\begin{enumerate}
%
%   \item {\em (initialization)} Initialize  $P_{mc}$ of size $q\times m\times m'$.   
%    \item {\em (permute)} and {\em (hash)} as above XXX
%    \item {\em (thinning)} Calculate sub-stream index $s=h_t(\omega)$.
%    
% 
%   \item {\em (signal updates)} Update  $P_{mc}$ for $i \in [q]$:
%   $$
%    P_{mc}[i, h_i\circ \sigma(\omega), s] = P_{mc}[i,h_i\circ \sigma(\omega), s] + v, 
%   $$
%   
%   \item {\em (identification)} Upon populating the signal arrays for some period of time,
%   \begin{enumerate}
%   \item  {\em (max-count)} Construct matrix $P_{max}$ of size $q\times m$, i.e., for $i\in[q]$ and
%   $j\in [m]$ set $ P_{max}[i, j] = \max_{s} P_{mc}[i,j,s]$
%   \item ({\em select})  Find $o_i = {\rm Argmax}_{j} P_{max}[i,j]$
%   \item ({\em decode})
%     Output $\omega^* = \sigma^{-1}(\overline{o_4 \ldots o_1})$.
%    \end{enumerate}
%\end{enumerate}
 
\floatname{algorithm}{Algorithm} 
\setcounter{algorithm}{1}
\begin{algorithm}[t]
\caption{Max-Count Hashing Pursuit}
\label{alg:max-count}
%\scriptsize
\begin{algorithmic}[1]
\REQUIRE Number heavy-hitters: $k$
\STATE [Start]  Initialize  $P_{mc}$ of size $q\times m\times m'$.
\STATE Perform steps 2 \& 3 of Algorithm~\ref{alg:shp}
\STATE [Thin] Calculate sub-stream index $s=h_t(\omega)$.
\STATE [Update]    Update  $P_{mc}$ for $i \in [q]$:
   $$
    P_{mc}[i, h_i\circ \sigma(\omega), s] = P_{mc}[i,h_i\circ \sigma(\omega), s] + v, 
   $$
 \vspace{-10pt}  
\IF {(Populated $P_{mc}$ for some  time window)}
 \IF{k==1}
	\STATE  Construct matrix $P_{max}$ of size $q\times m$, i.e.,
	 \STATE $P_{max}[i, j] = \max_{s} P_{mc}[i,j,s]$, \ $i\in[q]$, $j\in [m]$ 
	 \STATE Find $o_i = {\rm Argmax}_{j} P_{max}[i,j]$, \ $i\in[q]$ 
	 \STATE [decode]      Output $\omega^* = \sigma^{-1}(\overline{o_q \ldots o_1})$. 
 \ELSE
  \FOR{r=1 \TO k} 
	 \STATE $P_{max}[i, j] = \max_{s\in [m']} P_{mc}[i,j,s]$, \ $i \in [q]$, $j \in [m]$ 
	 \STATE Find $o_i = {\rm Argmax}_{j} P_{max}[i,j]$, \ $i\in[q]$ 
	 \STATE Find $\ell_i = {\rm Argmax}_{\ell\in[m']} P_{mc}[i,o_i,\ell]$, \ $i\in[q]$  
	 \STATE $P_{mc}[i, o_i, \ell_i]$ = 0, \ $i\in[q]$  {\#Exclude for next iteration }
	 \STATE [decode]      Output $\omega_r^* = \sigma^{-1}(\overline{o_q \ldots o_1})$. 
  \ENDFOR
 \ENDIF 
 \ENDIF   
\end{algorithmic}
\end{algorithm}

%{\bf Remark} 
%The precision of this method can be improved by considering several (statistically) independent hash functions and then pooling the results from 
%the corresponding sub-sampled stream for each hash.
 
 \subsection{Hash-thinned MJRTY Boyer-Moore}
 \label{sec:bm}
 
 Algorithm~\ref{alg:bm} is based on the \emph{Boyer-Moore} majority vote algorithm~\cite{boyer_moore1991} (not
 to be confused with the Boyer-Moore algorithm for string matching  used in
 signature-based detection tools like~\emph{Snort}),
 and the idea of stream thinning for creating sub-streams described above. The
 MJRTY Boyer-Moore algorithm can identify \emph{exactly} the element
 in stream $\cal{S}$ with the largest traffic assuming that its volume is at least 50\% of the total volume (i.e., there is a majority).
 It solves the problem in time linear in the length of the input sequence and constant memory.
 In reality, though, a single IP or flow is not usually responsible for such a high fraction of the total volume of the stream. Hence,
 identification accuracy of the plain Boyer-Moore algorithm could be lacking. To overcome this, we employ the Boyer-Moore
 idea on sub-streams of $\cal{S}$, constructed as described above. Thus, with, say, $m=256$ sub-streams 
 the `majority-threshold' for each sub-stream becomes $\frac{50}{m}$\% ($\approx 0.2$\%) of  stream $\cal S$ volume. In practice, as Fig.~\ref{fig:ident_accuracy_vol} illustrates, this makes  the Boyer-Moore-based algorithm to perform remarkably well, but one could use a higher $m$ to further thin stream $\cal{S}$, if needed.  
 
 We describe the original  Boyer-Moore algorithm with an analogy to the \emph{one-dimensional}
 random walk on the line of non-negative integers.  A variable $count$ is initialized to $0$ (i.e., the origin)
 and a  candidate variable \emph{cand} is reserved for use. Once a new IP arrives\footnote{We focus on single IPs. Everything we say here, though, apply on (src,dst) pairs,
and tuples of kind (src, sport, dst, sport) where sport/dport are the source and destination ports.},  
 we check to see if \emph{count} is 0. If it is, that IP is set to be the new candidate \emph{cand}
and  we move \emph{count} one step-up, i.e. $count=1$. Otherwise, if the IP is the same as \emph{cand}, then 
\emph{cand} remains unchanged and \emph{count} is incremented, and, if not, \emph{count} moves one step-down 
(decremented). We then proceed to the next IP and repeat the procedure. Provably, when all IPs are read,
\emph{cand} will hold  the one with majority, if majority exists.

We implemented a natural extension of the MJRTY Boyer-Moore algorithm to the case of byte and packet 
counts and applied it in parallel on the sub-streams resulting from hash-thinning. For each sub-stream, we also 
maintained an additional small and constant size data structure, used to estimate the signal volume.

\floatname{algorithm}{Algorithm} 
\setcounter{algorithm}{2} 
\begin{algorithm}[t]
\caption{Hash-thinned MJRTY Boyer-Moore}
\label{alg:bm}
%\scriptsize
\begin{algorithmic}[1]
\REQUIRE Number heavy-hitters: $k$, $k\le m$
\STATE [Start]  Initialize  $P_{bm}[i,j]=0$, \ $i\in[m]$, $j\in [m']$. 
\STATE Initialize  $count[i] = 0$,  $cand[i] = -1$, \ $i\in[m]$
\STATE Upon observing $(\omega, v)$, compute $\sigma (\omega)$.
\STATE [Thin] Calculate sub-stream index $s=h_t(\omega)$.
\STATE [Hash] Compute $j = h \circ \sigma(\omega)$, $h: [N] \to [m']$
\STATE [Update]
\IF{$cand[s] == -1$}
   \STATE $cand$[s] = $\omega$, $count$[s] = $v$, $P_{bm}$[s,j] = $v$
\ELSE
    \IF{$cand[s]==\omega$}
         \STATE $P_{bm}$[s,j] =  $P_{bm}$[s,j]  + $v$, $count$[s] =  $count$[s] + $v$
     \ELSE
          \IF{$count[s] > 0$}
               \STATE $count[s]$ = $count[s]$ - $v$
               \IF{$count[s] <0$}
                    \STATE $cand$[s] = $\omega$, $count$[s] = -$count$[s]   {\# reset candidate}
               \ENDIF
               \STATE $P_{bm}$[s,j] =  $P_{bm}$[s,j]  + $v$
          \ELSE 
             \STATE  $cand$[s] = $\omega$,  $count$[s] = $v$ {\# reset candidate}
             \STATE  $P_{bm}$[s,j] = $P_{bm}$[s,j]  + $v$ 
          \ENDIF   
    \ENDIF
\ENDIF
\IF {(Populated $P_{bm}$ for some  time window)}
     \STATE $P_{est} = \max_{j\in[m']} P_{bm}[i,j]$, \ $i\in[m]$
       \STATE Initialize ${\cal O} = \emptyset$
   \FOR{r=1 \TO k} 
        \STATE Find $o = {\rm Argmax}_{j\in[m] \setminus {\cal O}} P_{est}[j]$ 
        \STATE ${\cal O}  = {\cal O} \cup  o $  {\#Exclude for next iteration }
         \STATE Output $\omega_r^* = $cand$[o]$ 
  \ENDFOR
\ENDIF
%\STATE Sort based on array $P_{bm}$. Output heavy-hitter.
\end{algorithmic}
\end{algorithm}

\subsection{Applications to more complex signals}
\label{sec:ms}

 In this section we extend the randomized  {\em domain hashing} approach to the detection of more 
 complex anomalies. As discussed in Section~\ref{sec:streaming}, one common scenario could involve 
a ``persistent" user $\omega^*$ that accesses abnormally large
 number of ports.  Note that the overall volume or frequency of packets from user $\omega^*$ need not be abnormal. Therefore, such anomalous activity may be
 nearly impossible to detect from  algorithms that merely detect large volume hitters. 
 To address this problem with constant and small-size memory, 
 we propose next another algorithm, which combines the {\em randomized hashing} and
 {\em max-stable sketches}~\cite{4221749}.
 
 %The role of the max-stable sketch is to estimate the size of the set of {\em ports} with 
 %a given randomized octet value. This could have been done with other conventional hash--table techniques exactly, 
 %perhaps at the expense of maintaining a larger data structure and  higher computational cost. 
 %The {\em max-stable sketch} may be also easily modified to estimate more intricate characteristics of the signal such as 
 %dominance norms and it has the attractive analytic property of {\em max--linearity}, which allows us to seamlessly
 %combine estimates from different 
 %time points or different networks.
 
 These sketches exploit the {\em max-stability} property \eqref{e:max-stab} of the Fr\'echet distribution.
 Recall that a random variable $Z$ is $1$-Fr\'{e}chet if $\P\{Z \le x\} = \exp\{-c/x\}$, for $x>0$, and
 0, otherwise, for some scale $\sigma>0$. If $c=1$, then $Z$ is called  {\em standard $1$-Fr\'echet}.
 Let $Z, Z(1), \ldots, Z(n)$ be independent standard 1-Fr\'{e}chet random variables. Then, it is known~\cite{4221749} that 
 \begin{equation}\label{e:max-stab}
 \xi = \max_{1\le i \le n} Z(i) \,{\buildrel d \over =}\, nZ,
 \end{equation}
 where `${\buildrel d \over =}$' denotes equality in distribution.
Thus, $\xi$ is a 1-Fr\'{e}chet variable with scale coefficient $c=n$. One can easily express
the median of a 1-Fr\'{e}chet variable $Z$ with scale coefficient $n$ by considering that
$\P\{Z < \rm{med}(Z)\}$ = 1/2. By the definition of the 1-Fr\'{e}chet distribution and some algebra,
one obtains $\rm{med}(Z)=n/{\rm ln}2$.

The theory above can be employed to estimate the number of elements of a set.
Specifically, consider the problem of finding the number of {\em unique} destinations ports
contacted by a given host. When a stream element $(\omega, v)$ arrives, we generate a 1-Fr\'{e}chet
variable with seed that is a function of $v$. Thus,  if the same port arrives, the exact 
pseudo-random number $Z(v)$ is generated. As different ports arrive, we sequentially update 
our sketch by taking the maximum of the new 1-Fr\'{e}chet variable
and the sketch entry. Essentially, we are building one realization of the $\xi$ random variable 
described above.  The number of different terms $Z(v)$'s will correspond to the number of different ports.

Since, we want to take the median of $\xi$ to estimate the scale coefficient $n$,
we independently keep $L$ realizations of the above procedure (see Alg.\ \ref{alg:mshp}).
 
% \medskip
% \noindent{\bf Algorithm: (\emph{Max-Stable Hashing Pursuit})}  
% 
% \begin{enumerate}
%  \item (\emph{initialization}) Initialize array $P_{ms}$ of size $q\times m \times K$, i.e., set $P_{ms}[i,j,k]=0$,  for $k \in [K]$, $j \in [m]$, and $i\in[q]$.
%  \item {\em (permute)} Upon observing $(\omega, v)$, compute $\sigma (\omega)$.
%  \item {\em (pseudo-random number generation)} With \emph{random seed} $v$, and $k\in[K]$, generate $K$ independent 1--Fr\'echet random variables $Z_k(v)$.
%  \item {\em (hashing)} Compute $v_i = h_i\circ \sigma (\omega), i \in [q]$.
  
 % \item (\emph{signal updates})  Compute $P_{ms}[i, v_i, k ] = P_{ms}[i, v_i, k ] \vee Z_k(v)$, for $i\in[q]$, $k\in[K]$, 
 % where $\vee$ denotes pairwise maximum operation.
 %    \item {\em (identification)} Upon populating the signal arrays for some period of time,
 %  \begin{enumerate}
  % \item  {\em (max-count)} Construct matrix $P_{med}$ of size $q\times m$, i.e., for $i\in[q]$ and
  % $j\in [m]$ set $ P_{med}[i, j] = {\rm med}_{k} P_{mc}[i,j,k]$
  % \item  ({\em select})  Find $o_i = {\rm Argmax}_{j} P_{med}[i,j]$ for $i\in[q]$.
  % \item ({\em decode})
  %   Output $\omega^* := \sigma^{-1}(\overline{o_4 \ldots o_1})$.
  %  \end{enumerate}
  %\end{enumerate}
  
\begin{algorithm}[t]
\caption{Max-Stable Hashing Pursuit (MSHP)}
\label{alg:mshp}
%\scriptsize
\begin{algorithmic}[1]
\REQUIRE Size of max-sketch $L$
\STATE [Start]    Initialize array $P_{ms}$ of size $q\times m \times L$,
\STATE [Permute]  Upon observing $(\omega, v)$, compute $\sigma (\omega)$
\STATE [Hash]  Compute $v_i = h_i\circ \sigma (\omega), \ i \in [q]$.
\STATE [RNG]  With \emph{random seed} $v$, and $\ell\in[L]$, generate $L$ independent 1--Fr\'echet random variables $Z_\ell(v)$.
\STATE [Update]    Update $P_{ms}$ for $i \in [q]$, $\ell\in[L]$:
   $$
    P_{ms}[i, v_i, \ell ]  = \max \{P_{ms}[i, v_i, \ell ], Z_\ell(v) \}.
   $$
 \vspace{-10pt}  
\IF {(Populated $P_{ms}$ for some  time window)}
   \STATE Construct matrix $P_{med}$ of size $q\times m$, i.e.,
   \STATE $ P_{med}[i, j] = {\rm med}_{\ell} P_{mc}[i,j,\ell]$, \ $i\in[q]$ and $j\in [m]$ 
   \STATE Find $o_i = {\rm Argmax}_{j} P_{med}[i,j]$, \ $i\in[q]$ 
 \STATE [Decode]  Output $\omega^* = \sigma^{-1}(\overline{o_q \ldots o_1})$. 
 \ENDIF    
\end{algorithmic}
\end{algorithm}

 %The median $P_{ms}[\ell, h_\ell\circ\sigma(j),k],\ k=1,\ldots,K$ can be used to estimate the scale of these r.v.'s, which equals
 %$$
 %\#\{ {\rm ports},\ \mbox{ used by all keys $j$ with the same randomized $\ell$-th octet value } \}.
 %$$
 %\item (identify) A simple histogram of 
 %$$
%m_\ell(o):= {\rm med} ( P_{ms}[\ell, u, k],\ k=1,\ldots,K ),\ u \in \{1,\ldots,256\}
% $$
% will involve peaks for each $\ell=1,\ldots,4$. The octet value corresponding to these peaks will allow us to recover the IP of the culprit.

\section{On the accuracy of identification}
\label{sec:bounds}

 The proposed algorithms (except the hash-tinned MJRTY Boyer-Moore) are closely related to the general family 
 of sub-linear algorithms in the theoretical work of \cite{Porat:2012:STM:2095116.2095212}. Performance guarantees 
 for our algorithms can thus be established in a similar way as for the so-called Euclidean 
 {\em approximate sparse recovery systems} of  \cite{doi:10.1137/100816705}.  

To understand better the role of {\em collisions}, however, we adopt a different approach and provide 
performance guarantees under the following simplifying assumptions.  All  proofs are available in the 
Appendix.

\smallskip
\noindent{\bf Assumption 1.} {\em (exact $k$-sparcity)} The signal $f:\Omega\to V$ has precisely $k$ non-zero (non-empty) entries
at keys $\omega_i,\ i\in [k]$.

\smallskip
\noindent{\bf Assumption 2.} {\em (separation)} The magnitude functionals $L(i):= {\cal L}(f(\omega_i)),\ i\in [k]$ 
of the heavy-hitters are all different:
$
 L (1) >L(2) > \cdots >L(k)> 0. 
$
Define $\overline{L}(i) := \sum_{j=i+1}^k L(j)$ to be the cumulative magnitude of the bottom $(k-i)$ heavy hitters. By convention
set $\overline L(k) = 0.$

%\medskip
%Our analysis can be extended to the non-sparse case where the signal has additive ``noise" component, which is 
%suitably negligible, relative to the heavy hitters. 

% \subsection{Exact Recovery Guarantees}
 \smallskip
 \noindent{\bf Exact recovery guarantees.} Clearly, under Assumption 2, all our algorithms will identify 
 the top-$k$ heavy hitters {\em exactly}, provided that there are no hash-collisions. Let $p_k(r)$ denote the 
 probability that the top-$r$ heavy hitters are correctly identified for a $k$-sparse signal $f$. The following 
 results provide lower bounds on $p_k(r)$ under various conditions.

\begin{thm} \label{p:simple} Let $r\in[k]$ and suppose that $L(r)>\overline L(r)$. Then, for the Simple Hashing Pursuit algorithm:
\begin{eqnarray}\label{e:b-day}
 p_k(r) &\ge&  \left(1- \frac{r}{m}\right)^{(k-r)q} \prod_{i=1}^{r-1} \left(1-\frac{i}{m}\right)^q    \\
 &\ge& 1 - \frac{q r(k-1)}{2m}.\nonumber 
\end{eqnarray}
\end{thm}
\noindent{\em Note:} With $\Omega=\{0,1\}^{32}$ and the choice of hash functions as in Section~\ref{sec:hash-ident}, we have  $q=4$ and $m=256$. 
\HideProof{\begin{proof}
For each hash function $h_j,\ j\in [q]$, let $A_j(r)$ be the event that the top-$r$ heavy hitters are hashed into $r$ different bins,
{\em and} at the same time, the remaining $k-r$ hitters are hashed into the remaining $m-r$ bins. That is, the bins of
the top-$r$ hitters involve no collisions.  By the independence of $h_j(\omega_i),\ i\in [k]$, we have
\begin{equation}\label{e:simple-1}
\P(A_j(r)) = \frac{(m-1)}{m} \cdots \frac{(m-r+1)}{m} \times \left (\frac{m-r}{m}\right)^{k-r}.
\end{equation}
If the event $\cap_{j\in[q]} A_j(r)$ occurs, then the top-$r$ heavy hitters will be correctly identified. Thus, the
independence of the events $A_j(r)$ in $j\in [q]$ implies that 
$$
p_r(k) \ge \P(\cap_{j\in [q]} A_j(r)) = \P(A_j(r))^q, 
$$
which by \eqref{e:b-day} yields the first bound. The second bound follows from 
the {\em product comparison} inequality
$$
|\prod_{i=1}^k a_i - \prod_{i=1}^k b_i |\le \sum_{i=1}^k |a_i - b_i|,
$$
valid for all $a_i,b_i\in[-1,1]$. Indeed, setting $a_i=1$ and $b_i = (m-i\wedge r)/m$, we get
\begin{eqnarray*}
1- \P(A_j(r))^q &\le& q \sum_{i=1}^{r-1}\frac{i}{m} + (k-r)\frac{r}{m} \\
&=& q\frac{ r(r-1) + (k-r)r}{2m},
\end{eqnarray*}
which gives the second bound.
\end{proof}}

\begin{rem}[the birthday problem] For $r=k$ and $q=1$, the first bound in \eqref{e:b-day} can be interpreted as
the probability that a class of $k$ students have all different birthdays, if the year has
$m$ days. It is well-known that this probability decays rather quickly as $k$ grows.
\end{rem}
The max-count algorithm addresses this curse of the {\em birthday problem} by effectively increasing the value 
$m$.

\begin{cor}
\label{cor:1}
 Under the Assumptions of Proposition \ref{p:simple}, for the Max-count Hashing Algorithm,
the bounds in \eqref{e:b-day} apply with $m$ replaced by $m\times m'$.
\end{cor}
\HideProof{\begin{proof}
The result follows by observing that hash-thinning with an independent uniform 
hash function $h_t$ taking $m'$ values leads to $m\times m'$ bins in \eqref{e:simple-1}.  
\end{proof}}

Since the Boyer--Moore algorithm uses one hash function (as opposed to $q$), 
we obtain the following.

\begin{cor}  Under the Assumptions of Proposition \ref{p:simple}, for the hash-thinned MJRTY Boyer-Moore
Algorithm, the bounds in \eqref{e:b-day} apply with $q:=1$.
\end{cor}

\smallskip
\noindent
%\subsection{Bounds on the rate of identification}
{\bf Bounds on the rate of identification.} The {\em exact} identification of the top-$r$ hitters is difficult 
as the above performance bounds indicate since a few collisions may lead to misspecification. 
Even in the presence of collisions, however, a relatively large proportion of the hitters is identified 
in practice (see e.g.\ Fig.~\ref{fig:ident_accuracy_vol} below).  This suggests examining the 
{\em rate of identification} quantity:
\begin{equation}\label{e:e_r}
 e_{rt}(k) = \E N(k)/k,
\end{equation}
where $N(k)$ denotes the number of correctly identified heavy hitters among the top $k$. That is, $e_{rt}(k)$
gives the average proportion of identified top hitters.

The combinatorial analysis needed to establish bounds on $e_{rt}(k)$ for the
Simple Hashing and Max-count algorithms is rather involved and delegated to a follow-up paper. Under certain
conditions, however,  an appealing closed-form expression for $e_{rt}(k)$ is available for the hash-thinned MJRTY
Boyer-More algorithm. The conditions may be relaxed with the help of technical probabilistic analysis, which merits 
a separate investigation.

\begin{thm} Suppose that $L(i) > \overline{L}(i),$ for all $i\in[k]$. Then, under Assumptions 1 \& 2, for the hash-thinned MJRTY
Boyer-Moore Algorithm, we have the exact expression
\begin{eqnarray}
\label{e:bm-expected}
e_{rt}(k) &=& \frac{m}{k} \left(1- \left(1-\frac{1}{m}\right)^k \right)\\
&=& \frac{m}{k} \left( 1-e^{-k/m}(1+O({1}/{m})) \right).\nonumber
\end{eqnarray}
%see \url{http://www.math.uah.edu/stat/urn/Birthday.html}
\end{thm}
\HideProof{
\begin{proof} The assumption $L(i)> \overline L(i)$ guarantees that 
$N(k)$ equals the number of distinct values in the set of hashes $\{h(\omega_1),\cdots,h(\omega_k)\}$.
Let $Y_i = 1$ if bin $i$ is occupied and $0$ otherwise, for $i\in[m].$ Observe that
\begin{equation}\label{e:bm-expected-1}
\E N(k) = \sum_{i=1}^m \E Y_i = m \E Y_1,
\end{equation}
by exchangeability. Note, however, that 
$$
\E Y_1 = 1 - \P ( h(\omega_j) \not = 1,\ j\in[k]) = 1 - \left(1-\frac{1}{m} \right)^k,
$$
since $h(\omega_i),\ i\in[k]$ are independent and Uniform$([m])$.
This, in view of \eqref{e:e_r} and \eqref{e:bm-expected-1}, implies 
the first relation in \eqref{e:bm-expected}. The second follows from the standard
approximation $(1-1/m)^m \approx e^{-1}.$
\end{proof}}

\begin{figure}
         \vspace{-120pt}
        \centering
        \includegraphics[width=0.75\textwidth]{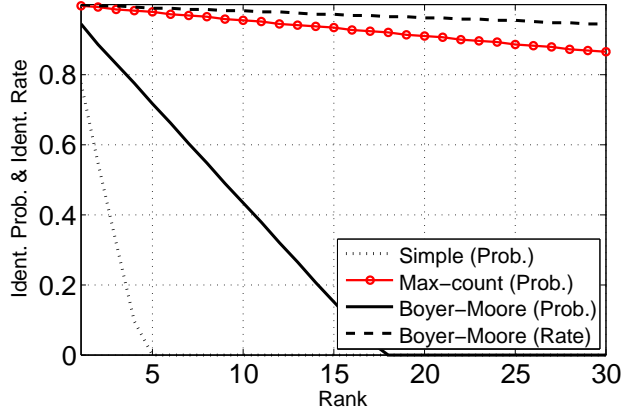}
         \vspace{-130pt}
        \caption{Performance bounds given the Assumptions of Section~\ref{sec:bounds}.}
        \label{fig:bounds}
 \end{figure}

\section{Performance Evaluation}

Next, we evaluate our algorithms with \emph{Netflow} traffic
traces provided from the academic ISP Merit Network. The traces are collected
at a large edge router. In the hourly-long trace we investigate,
our Netflow stream has $\approx360K$ unique source addresses, $\approx360K$
unique destinations, and consists of $\approx4.7$ million flow records.
The total volume is $103$ gigabytes and $132$ million packets.
We demonstrate results using Python software implementations
of our algorithms.

\begin{figure*}        
        \begin{subfigure}[b]{0.24\textwidth}
                \includegraphics[width=\textwidth]{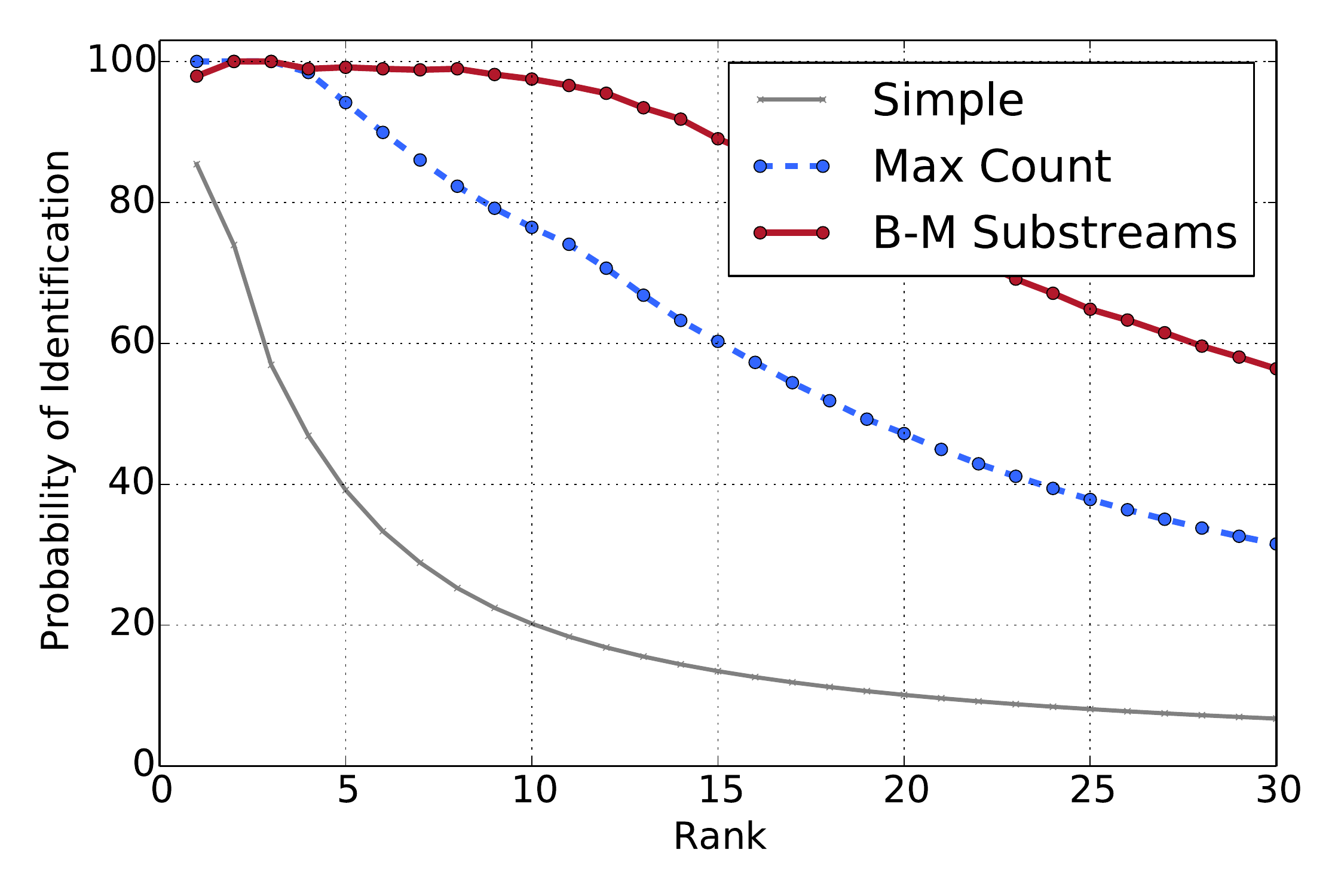}
                \caption{\footnotesize Frequency, ident. rate}
                \label{fig:freq_set100}
        \end{subfigure}
        \begin{subfigure}[b]{0.24\textwidth}
                \includegraphics[width=\textwidth]{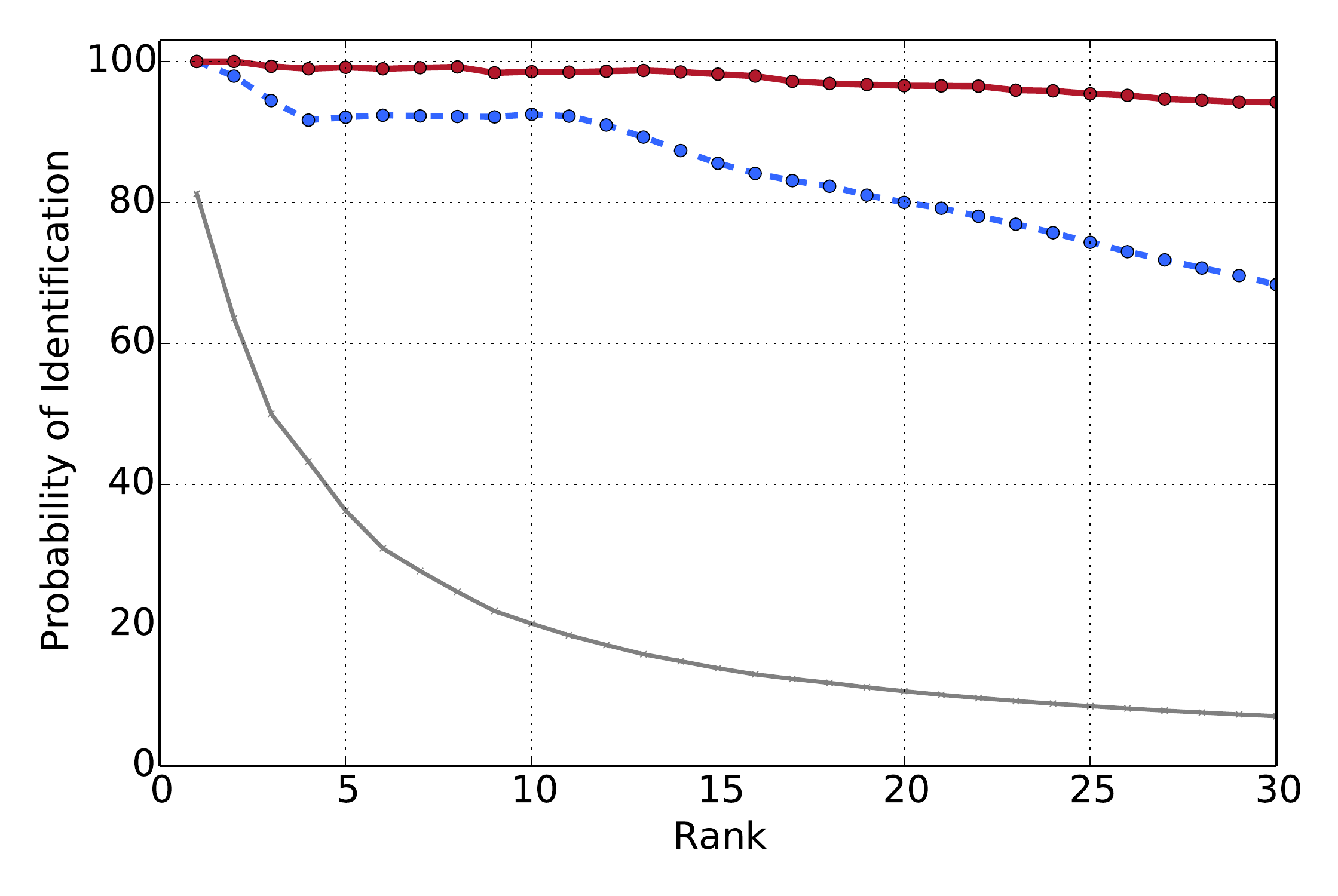}
                \caption{\footnotesize Payload, ident. rate}
                \label{fig:payload_set100}
        \end{subfigure}
        \begin{subfigure}[b]{0.24\textwidth}
                \includegraphics[width=\textwidth]{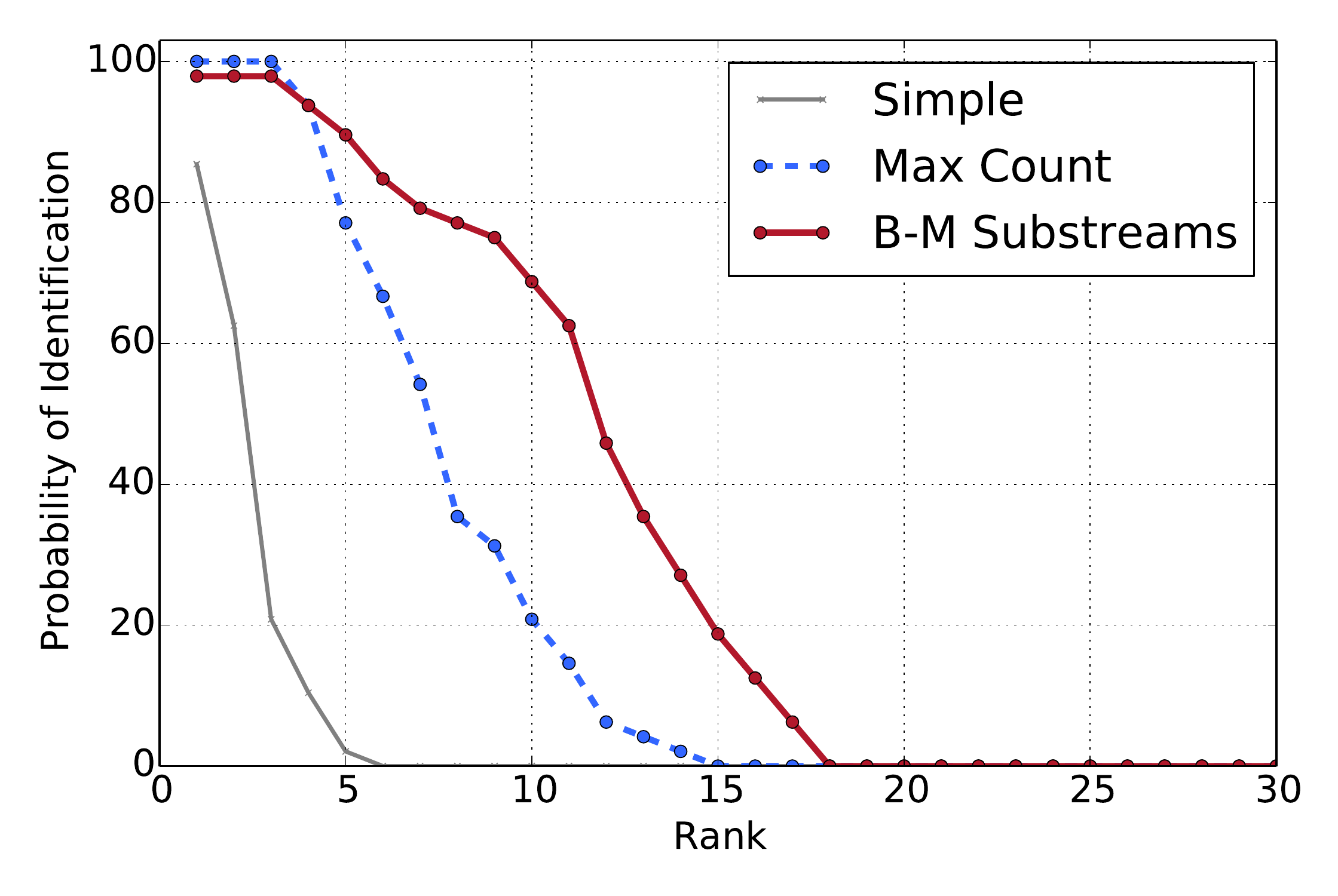}
                \caption{\footnotesize Frequency, exact metric}
                \label{fig:freq_binary100}
        \end{subfigure}
        \begin{subfigure}[b]{0.24\textwidth}
                \includegraphics[width=\textwidth]{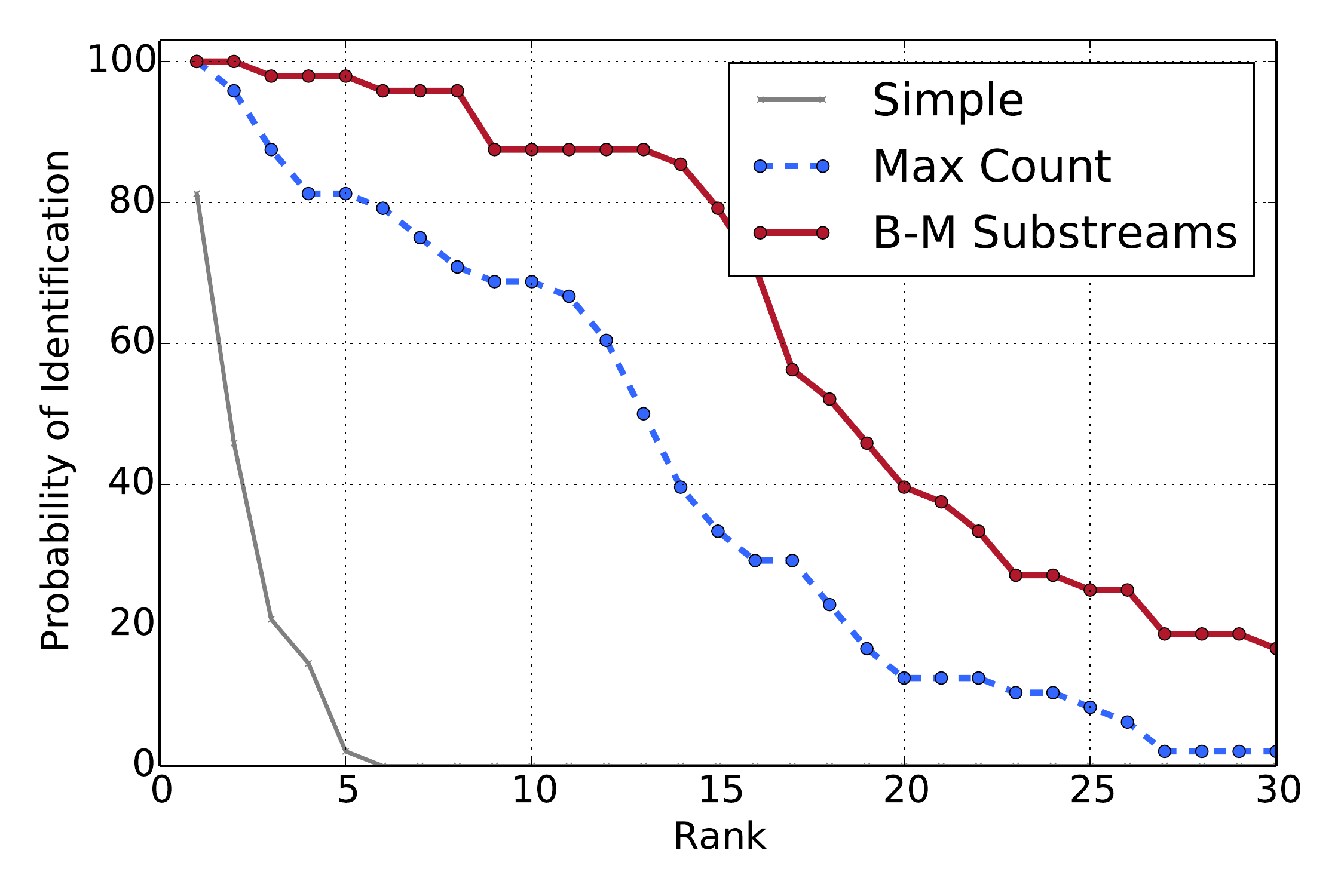}
                \caption{\footnotesize Payload, exact metric}
                \label{fig:payload_binary100}
        \end{subfigure}
        \vspace{-5pt}
        \caption{\footnotesize Identification accuracy results (window=100K).}
                \label{fig:ident_accuracy_vol}
\end{figure*}

%\subsection{Accuracy performance analysis}
\smallskip
\noindent{\bf Accuracy performance analysis.} Fig.~\ref{fig:ident_accuracy_vol} shows the identification accuracy of 
Algorithms~\ref{alg:shp}--\ref{alg:bm}, compared to the ground truth (obtained by counting using a dynamic hash 
array, and then sorting). We search for heavy hitters both in byte volume, as well as frequency of occurrence. We 
use two metrics to evaluate our methods: {\em (i)} identification rate and {\it (ii)} exact recovery. 
The first gives the expected proportion of identified heavy hitters among the top-$k$, while the second one is much 
more strict and measures the probability that all top-$k$ hitters are correctly recovered. 

\begin{figure*}
        \vspace{-12pt}
        \centering
        \begin{subfigure}[b]{0.3\textwidth}
                \includegraphics[width=\textwidth]{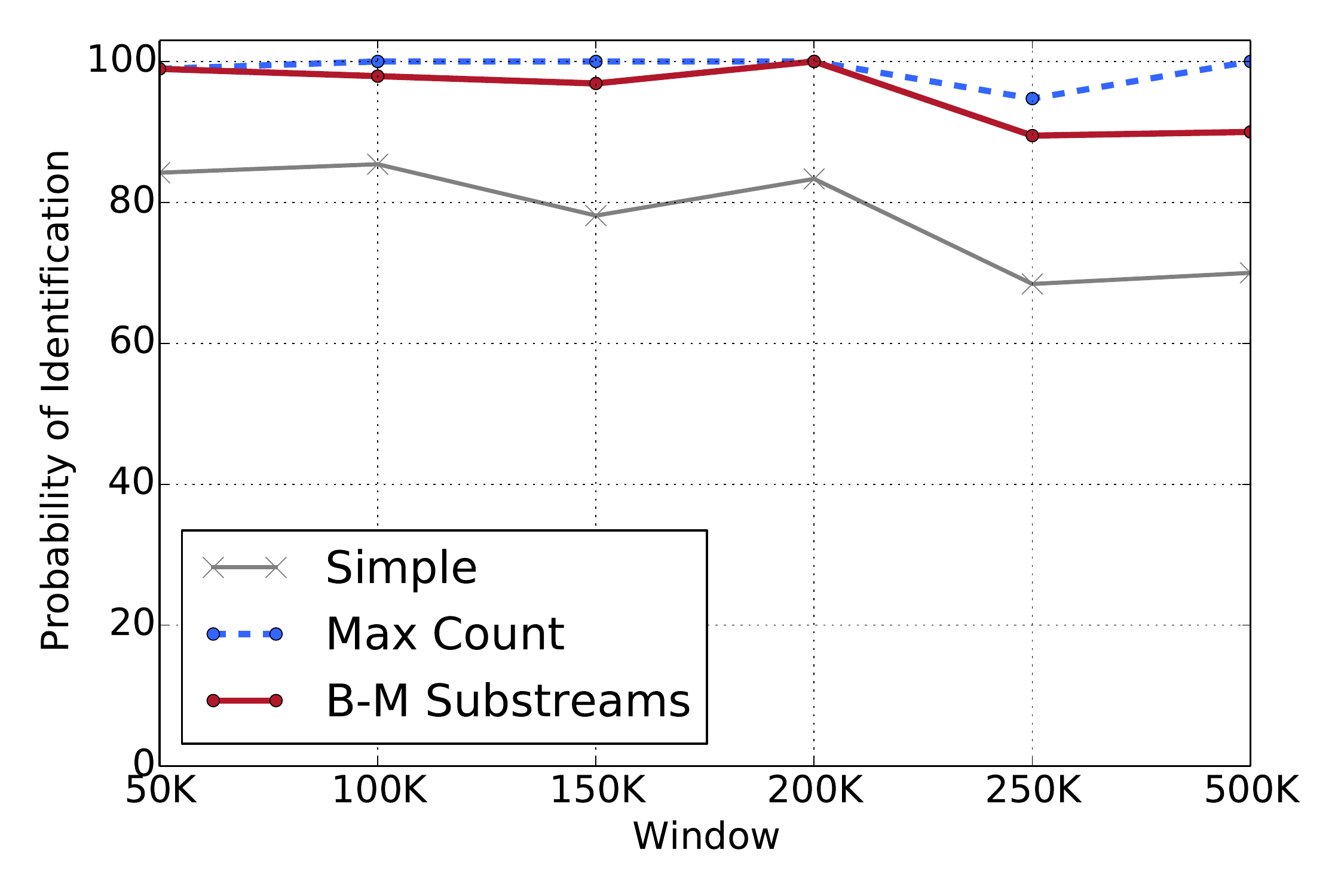}
                \caption{\footnotesize Frequency, top heavy-hitter}
                \label{fig:win:1}
        \end{subfigure}
        \begin{subfigure}[b]{0.3\textwidth}
                \includegraphics[width=\textwidth]{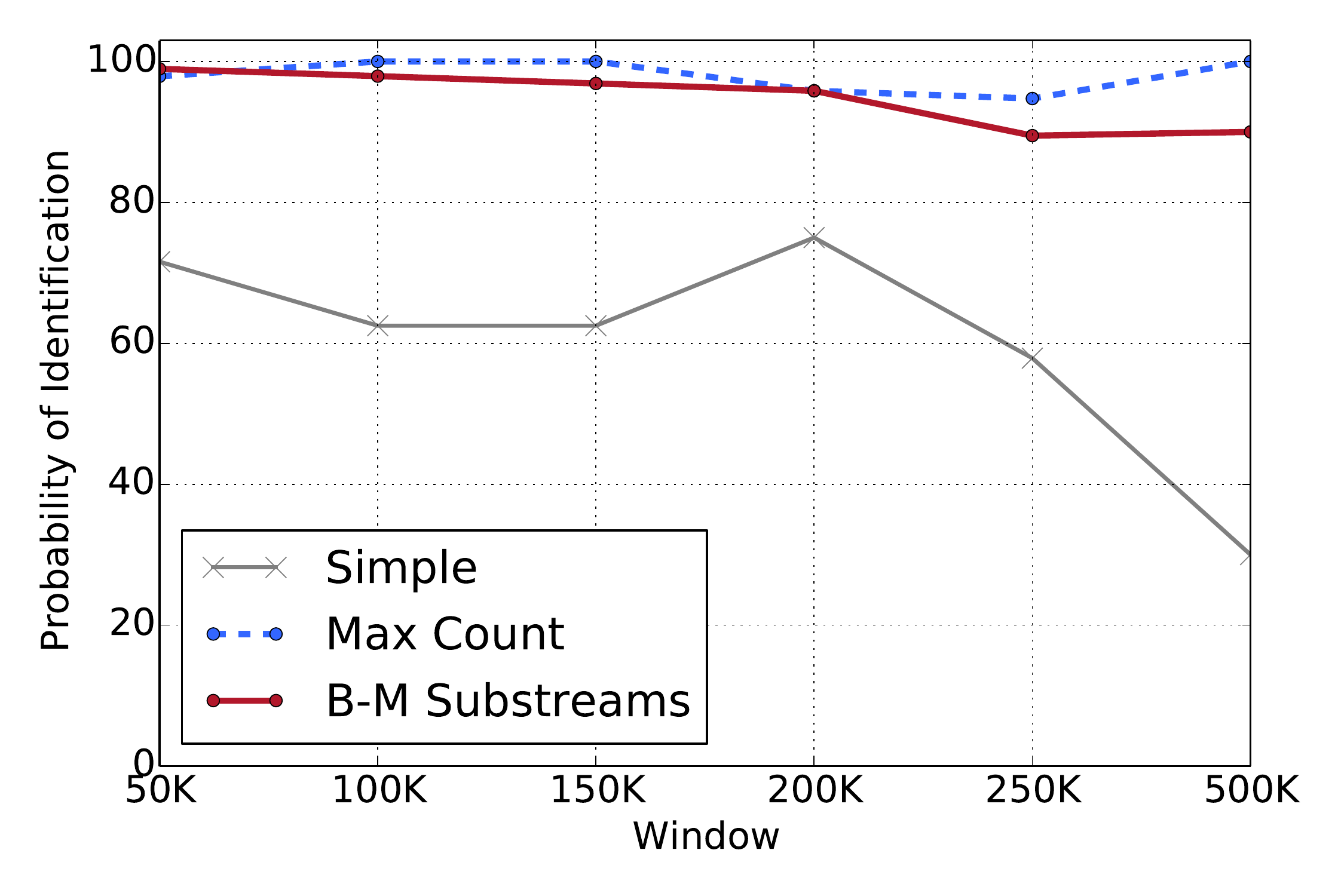}
                \caption{\footnotesize Frequency top-2 hitters}
                \label{fig:win:2}
        \end{subfigure}       
        \begin{subfigure}[b]{0.3\textwidth}
                \includegraphics[width=\textwidth]{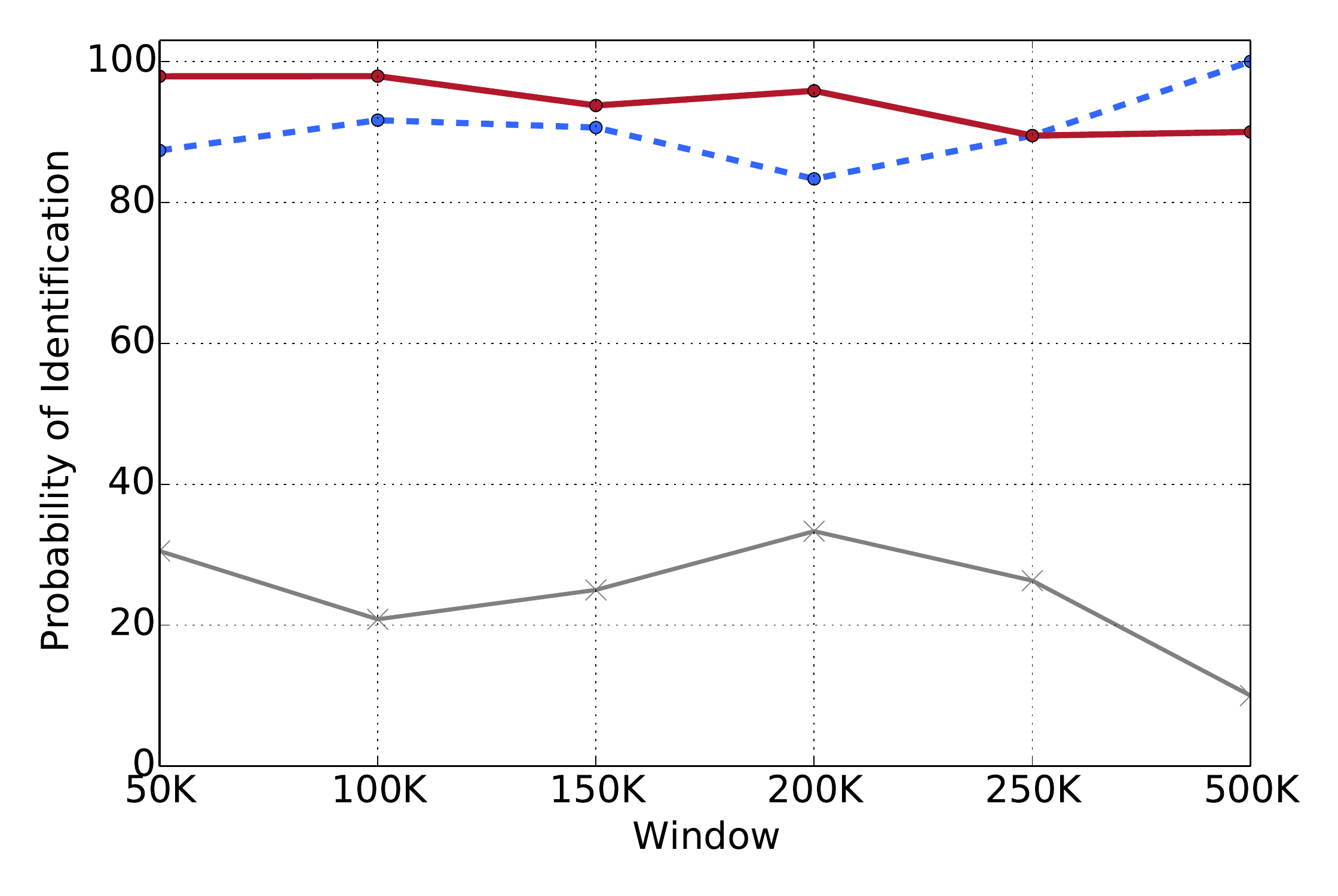}
                \caption{\footnotesize Frequency top-3 hitters}
                \label{fig:win:3}
        \end{subfigure}
        \caption{\footnotesize Identification accuracy (using the \emph{exact} criterion)  with varying window.}
        \label{fig:window_size_vary}
\end{figure*}

As Fig.~\ref{fig:ident_accuracy_vol} depicts, hash-thinned Boyer-Moore  and Max-Count perform remarkably well. 
The window of records we used is $100$K (i.e., we report the culprits every $100$K netflow records), which corresponds (for the stream studied)
on the RTS to slots of $~1.2$ minutes. The performance of the SHP
deteriorates as the number $k$ of top heavy hitters sought increases
which is an expected outcome of the \emph{birthday problem} discussed above.
Therefore, SHP is well-suited when stringent memory constraints are imposed,
and when only the top-hitter is of interest.
Fig.~\ref{fig:window_size_vary} provides a sensitivity analysis w.r.t the window size
that  one could perform to find the optimal monitoring window size.

%Note that in the RTS scale of
%interest, the Simple method could have high accuracy as well. Fig.~\ref{fig:window_size_vary}
%shows how the performance varies as the time ``window" increases. Clearly, Simple Hash-Count
%has a much better performance for smaller reporting intervals (2-3 minutes). 

Table~\ref{tab:est:accuracy:value} shows the estimation accuracy of our methods. 
We juxtapose the actual traffic volume of the heavy-hitter versus an estimate obtain from our sketches.
For Algorithms~\ref{alg:shp} \&~\ref{alg:max-count},   we estimate 
the volume by taking the average of the 4 octets
of the $m-$ary hash arrays. For thinned Boyer-Moore,
we report the value of array $P_{bm}$. Indeed,  Algorithm~\ref{alg:shp}'s 
collisions are inevitable, but for the other two, collisions are spread nicely, and
the estimated signal closely approximates  the true value (here we round to closest digit). More sophisticated estimates  (e.g., see~\cite{Krishnamurthy:2003:SCD:948205.948236} that uses a median-based estimate
to boost confidence) can also be considered.

%\begin{figure}[t]
%        \centering
%         \includegraphics[width=0.45\textwidth]{./figures/estimation_payload_accuracy_CHI600d1.pdf}
%        \caption{Estimation accuracy. We plot the accuracy of the estimate given by our algorithms
%        for top heavy-hitter volume (in bytes) compared to the exact value. Window size used is 500K.}
%        \label{fig:est:accuracy:value}
%\end{figure}

\begin{table}[t]
        \caption{\footnotesize Estimation accuracy (\%) of proposed algorithms
        for \emph{top} hitter  (in bytes) compared to the exact value (window size = 500K).}
        \label{tab:est:accuracy:value}
  \centering      
 \scriptsize       
\begin{tabular}{|l|cccccccl|}
\hline
{\bf Time} & {\bf 1}     & {\bf 2}     & {\bf 3}     & {\bf 4}     & {\bf 5}     & {\bf 6}     & {\bf 7}     & {\bf 8}       \\
\hline
\hline
Simple                        & 88  & 88  & 88  & 87  & 93  & 92  & 91  & 62 \\
Max-Count                     & 100 & 100 & 100 & 100 & 100 & 100 & 100 & 100  \\
Boyer-Moore                   & 100 & 100 & 100 & 100 & 100 & 100 & 100 & 100  \\
\hline
\end{tabular}
\end{table}

Table~\ref{tab:mem} illustrates the small memory requirements of our methods. The Max-Count and
Boyer-Moore-based algorithms consume considerably less than 1MB of fast memory, which remains constant. On the other hand, the na\"ive approach
of using a dynamically varying hash-table, has increasing memory requirement. 
To showcase the memory increase the algorithm was run against a stream with an hourly total byte volume of $1328$ gigabytes.
As demonstrated
in Table~\ref{tab:mem}, as the monitoring window grows (or equivalently, as the amount of traffic increases),
memory increases exponentially.

\begin{table}[t]
\caption{\footnotesize Memory Utilization (in MB)}
\label{tab:mem}
\centering
\begin{tabular}{|l| c c c c |}
\hline
{\bf Window Size ($\times$ 50K)}  & {\bf 1}    & {\bf 2--7}  & {\bf 8--19} & {\bf 20}     \\
\hline
\hline
Exact                                                 & .8 & 3  & 6  & 13  \\
Boyer-Moore ($m$=$m'$=256)                                                & .5 & .5 & .5 & .5   \\
Max-Count     ($q$=4, $m$=256, $m'$=50)                                       & .4 & .4 & .4 & .4  \\
\hline                                                      
\end{tabular}
\end{table}

\smallskip
\noindent{\bf Max-Stable sketch case studies.}
We illustrate Algorithm~\ref{alg:mshp}'s accuracy by looking at set-valued signals, tailored for detecting: 
(i) host scanning, (ii) port scanning, and (iii) a signal based on \emph{time to live} (TTL) values. 
%Note that, no filtering on 
%the hosts is applied here. For example, a heavy-hitter identified for case (i) could be, say, a benign server (as a source) that ``talks" with multiple
%clients (destinations) or other normal situations like this. However, for the purposes of algorithmic evaluation, 
%these considerations could be ignored.

\begin{figure*}[t]
        \centering
        \begin{subfigure}[b]{0.3\textwidth}
              \includegraphics[width=1.0\textwidth]{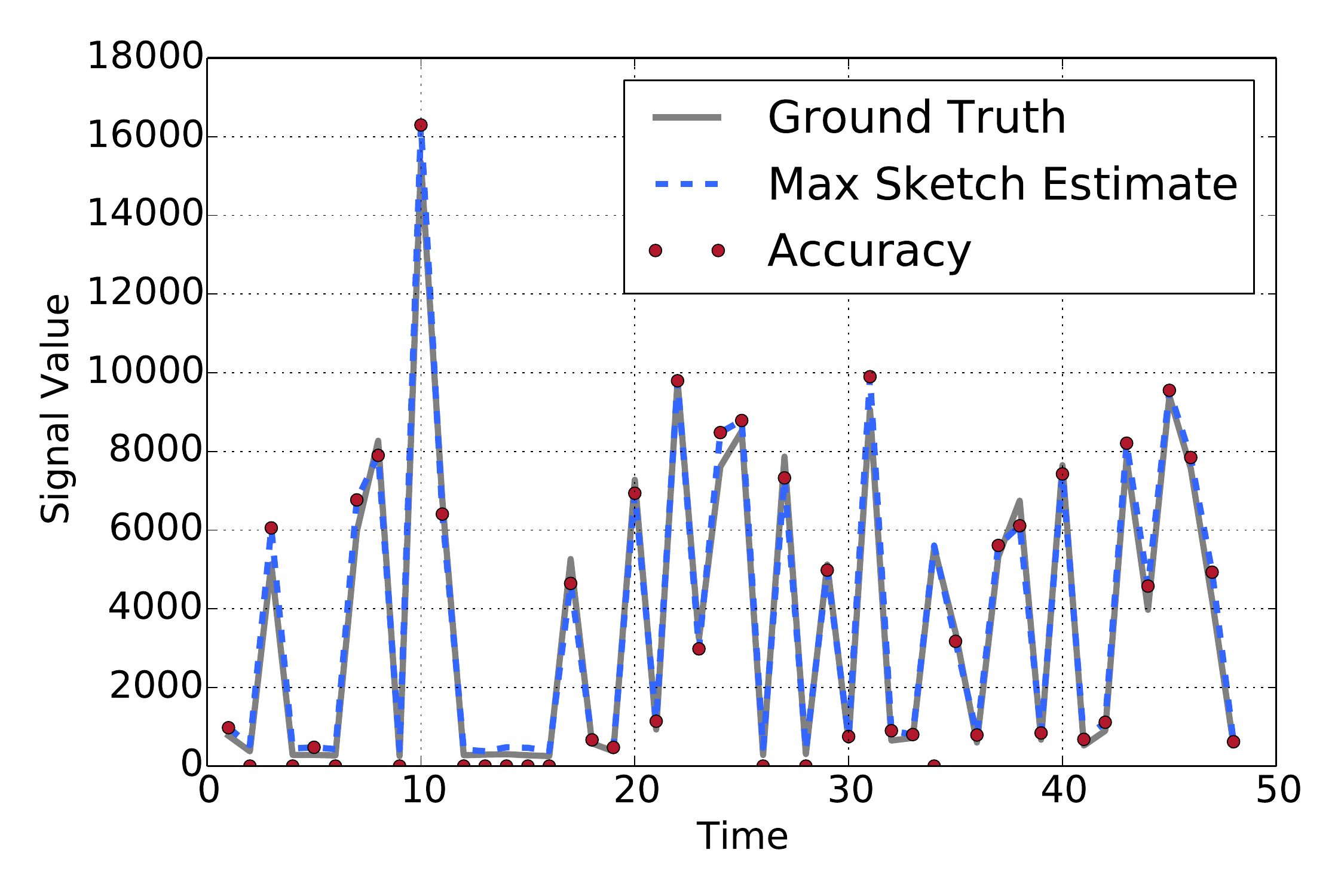}
                \caption{\footnotesize Host scanning}
                \label{fig:max_sketch_eval}
        \end{subfigure}
        \begin{subfigure}[b]{0.3\textwidth}
                \includegraphics[width=1.0\textwidth]{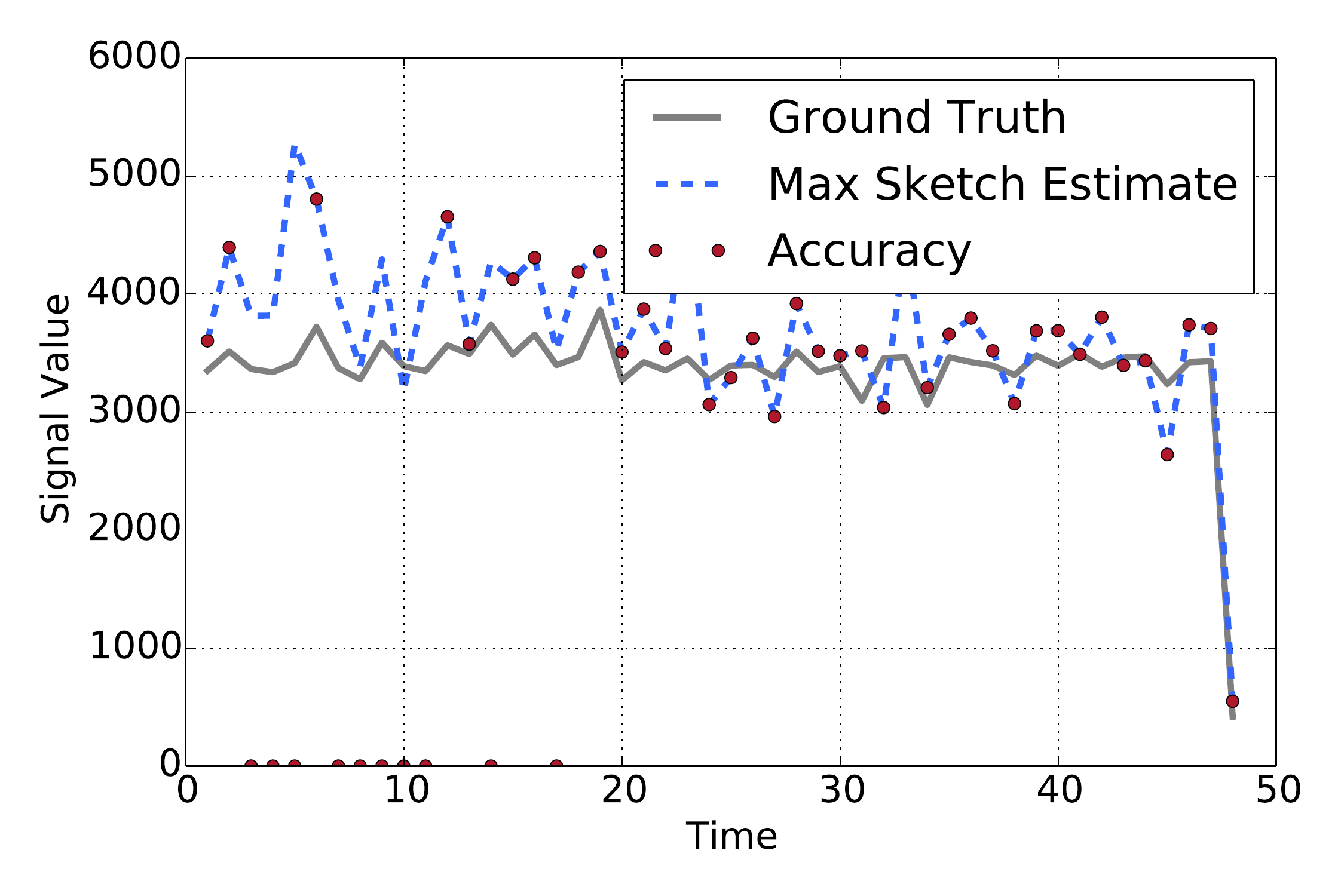}
                \caption{\footnotesize Port scanning}
                \label{fig:max_sketch_eval_ports}
        \end{subfigure}                
        \begin{subfigure}[b]{0.3\textwidth}
                \includegraphics[width=1.0\textwidth]{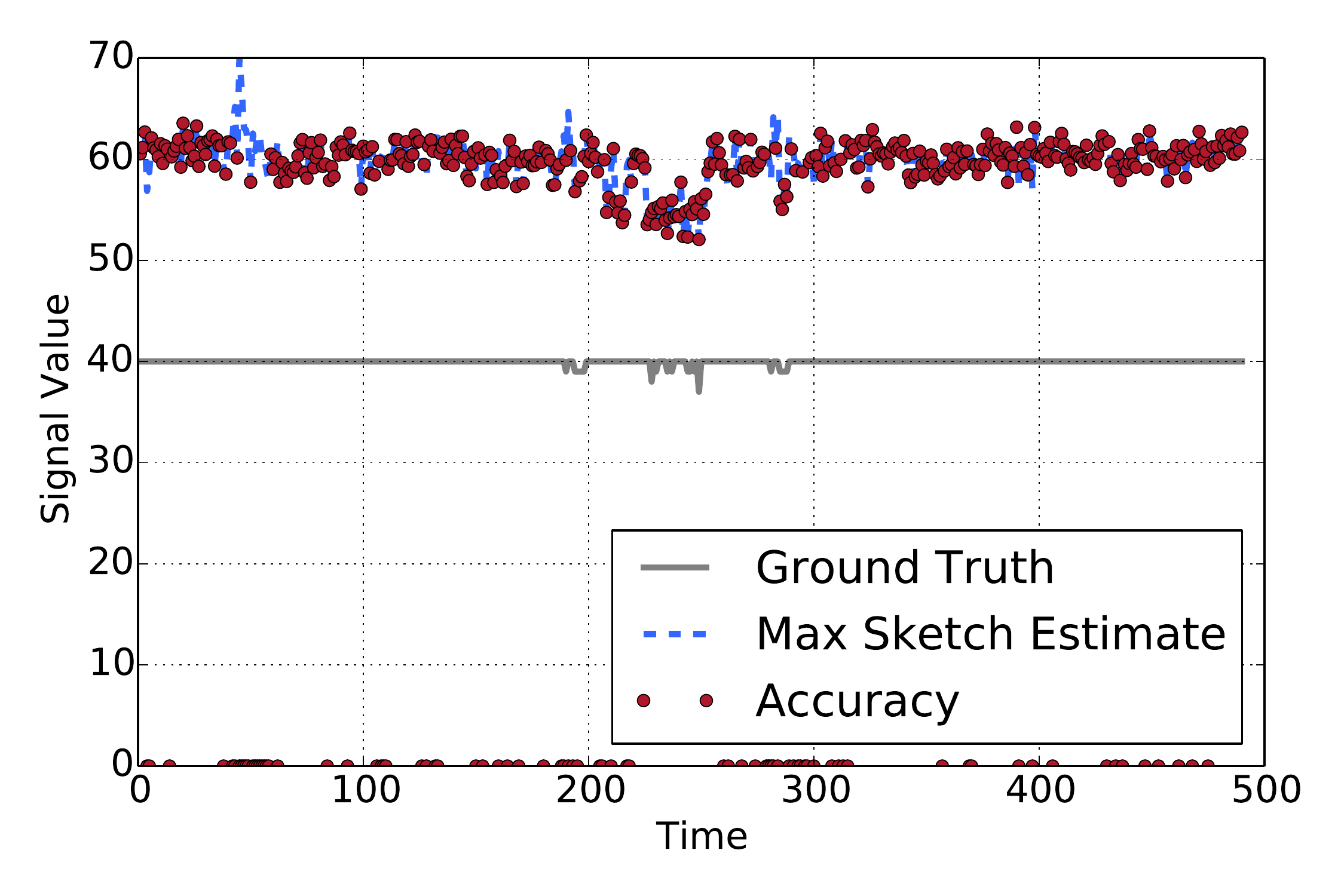}
                \caption{\footnotesize IP spoofing}
                \label{fig:max_sketch_eval_ports}
        \end{subfigure}
         \caption{\footnotesize Evaluation of Algorithm~\ref{alg:mshp} on the Netflow stream (window=100K).}
          \label{fig:max_stable_case_studies}
\end{figure*}

Fig.~\ref{fig:max_sketch_eval} shows the accuracy on {\em host scanning}, 
where the goal is to identify the source IPs that contact large number of destination IPs.  
A filled-circle on the dotted blue line indicates the occurrences we identify 
exactly the heavy-hitter IP. The solid gray line shows the exact number of different hosts that the malicious scanner has contacted. The dotted-blue
line shows our algorithm's estimate for the set cardinality obtained by simply averaging, as described above.
The identification accuracy
is very high, and the misses occur when the set-size is indeed quite small. 

In Fig.~\ref{fig:max_sketch_eval_ports}, our signal updates involve the set of ports that a given source IP interacts with.
Hence, this could be used to identify port scanners. While the accuracy is still quite high,
we observe a few mis-identifications. This is because the number of unique ports ($2^{16}$) is much smaller than the unique IPs ($2^{32}$).
As a result, collisions in our hash-arrays matter more in this case; background ``noise" from
non-malicious IPs could alter the ranking of even one of the buckets selected on the \emph{identification} step, and result in mis-identification.

In Fig.~\ref{fig:max_sketch_eval_ports}, we apply the max-stable sketch on Darknet data (known also as Internet background radiation)~\cite{wustrow:2010:radiation}. 
The dataset available does not include Netflow records, but instead consists of  packets captured at a network interface
(this 1-hour long trace  has $50$ million packets). Darknets are composed of traffic destined at unallocated address spaces 
(i.e., dark spaces). It is therefore directly
associated with malicious acts or misconfigurations.  Algorithm~\ref{alg:mshp} is used to identify packets whose source IP has been \emph{spoofed}.
To achieve this we look at the TTL values of each packet. For the problem at hand,  a heavy-hitter is a source IP whose set of unique TTL
values in the monitoring window is large -- an indication of spoofing~\cite{Beverly:2005:SPI:1251282.1251290}. 
The evaluation results illustrate that the culprit is correctly identified in around 84\% of the cases, % (410 out of 490)
and the result for deteriorating performance is again due to the small cardinality of the unique set of TTL values (which is $2^8$ in theory, but
in practice less unique TTL values are encountered).

\section{Conclusions}

 We presented a family of algorithms that are well-suited
 for online implementation in fast network streams. In addition,
 our framework can be employed to find heavy-hitters on
 a variety of signals, including complex ones that involve operations with sets.
 Further, our algorithms are amenable for distributed implementation;
in a decentralized setting, the proposed
 data structures could be constructed at various
observation points (i.e., the switch/router)
and then transmitted for aggregation at centralized decision centers
due to two important properties: (i)  they are small and constant in size and, hence,
can be efficiently emitted over the network to a centralized location, (ii)
they can be linearly combined/aggregated  by the centralized worker and reduced to
a single sketch object that can be utilized to yield the final culprits. 

%This distributed functionality is a very important consideration,
%having in mind the ever-increasing Internet traffic volume.
%Centralized-schemes with  monitoring appliances for DDoS detection (such as Arbor Networks Peakflow)
%being fed with a stream of network flows (usually sampled, due to high-speed rates)
%would not be able to scale as traffic grows.
%To deal with the ever-increasing traffic network operators would either have
%to increase the sampling rate or install more monitoring appliances. Both solutions are highly
%likely to introduce detection inaccuracy. A higher sampling rate  may introduce identification
%problems with ``small flows" (i.e., flows with small packet and byte volume) that are involved
%in ``lightweight" malicious activities such as horizontal host and port scanning.  On the other hand,
%additional monitoring appliances that work ``orthogonal" among themselves  (each one receiving streams from \emph{different} network nodes)
%could affect identification/detection accuracy for distributed attacks. For example,
%one can envision a scenario of distributed DoS traffic (such as the well-known NTP DDoS attacks emerging at the 
%onset of 2014) in which the traffic monitored at each individual monitoring station falls below a ``safety" threshold,
%but in reality the aggregate traffic volume entering the ISP's network is well above
%the threshold that should trigger an alert.

\appendix

\section{Proofs}

\subsection{Exact recovery guarantees}

The proof of Theorem~\ref{p:simple} follows.

\begin{proof}
For each hash function $h_j,\ j\in [q]$, let $A_j(r)$ be the event that the top-$r$ heavy hitters are hashed into $r$ different bins,
{\em and} at the same time, the remaining $k-r$ hitters are hashed into the remaining $m-r$ bins. That is, the bins of
the top-$r$ hitters involve no collisions.  By the independence of $h_j(\omega_i),\ i\in [k]$, we have
\begin{equation}\label{e:simple-1}
\P(A_j(r)) = \frac{(m-1)}{m} \cdots \frac{(m-r+1)}{m} \times \left (\frac{m-r}{m}\right)^{k-r}.
\end{equation}
If the event $\cap_{j\in[q]} A_j(r)$ occurs, then the top-$r$ heavy hitters will be correctly identified. Thus, the
independence of the events $A_j(r)$ in $j\in [q]$ implies that 
$$
p_k(r) \ge \P(\cap_{j\in [q]} A_j(r)) = \P(A_j(r))^q, 
$$
which by \eqref{e:b-day} yields the first bound. The second bound follows from 
the {\em product comparison} inequality
$$
|\prod_{i=1}^k a_i - \prod_{i=1}^k b_i |\le \sum_{i=1}^k |a_i - b_i|,
$$
valid for all $a_i,b_i\in[-1,1]$. Indeed, setting $a_i=1$ and $b_i = (m-i\wedge r)/m$, we get
\begin{eqnarray*}
1- \P(A_j(r))^q &\le& q \sum_{i=1}^{r-1}\frac{i}{m} + (k-r)\frac{r}{m} \\
&=& q\frac{ r(r-1) + (k-r)r}{2m},
\end{eqnarray*}
which gives the second bound.
\end{proof}

For Corollary~\ref{cor:1}, we have:

\begin{proof}
The result follows by observing that hash-thinning with an independent uniform 
hash function $h^{(s)}$ taking $m$ values leads to $m^2$ bins in \eqref{e:simple-1}.  
\end{proof}

\subsection{Bounds on the rate of identification}

The proof of Theorem~\ref{e:bm-expected} is as follows.

\begin{proof} The assumption $L(i)> \overline L(i)$ guarantees that 
$N(k)$ equals the number of distinct values in the set of hashes $\{h(\omega_1),\cdots,h(\omega_k)\}$.
Let $Y_i = 1$ if bin $i$ is occupied and $0$ otherwise, for $i\in[m].$ Observe that
\begin{equation}\label{e:bm-expected-1}
\E N(k) = \sum_{i=1}^m \E Y_i = m \E Y_1,
\end{equation}
by exchangeability. Note, however, that 
$$
\E Y_1 = 1 - \P ( h(\omega_j) \not = 1,\ j\in[k]) = 1 - \left(1-\frac{1}{m} \right)^k,
$$
since $h(\omega_i),\ i\in[k]$ are independent and Uniform$([m])$.

This, in view of \eqref{e:e_r} and \eqref{e:bm-expected-1}, implies 
the first relation in \eqref{e:bm-expected}. The second follows from the standard
approximation $(1-1/m)^m \approx e^{-1}.$
\end{proof}

\small
\balance
\bibliographystyle{IEEEtran}
%\bibliography{IEEEabrv,paper,detection}
\bibliography{detection}
\end{document}